\documentclass{aastex631}
\usepackage{graphicx}
\usepackage{amsmath}

\usepackage{color,soul}

\shorttitle{Ensemble Variability of AGN}
\shortauthors{Chanchaiworawit \& Sarajedini}

\begin{document}

\title{Ensemble Variability Properties of Active Galactic Nuclei in the SDSS DR17}

\author[0000-0002-9650-4371]{Krittapas Chanchaiworawit}
\affiliation{Department of Physics, Florida Atlantic University, \\
777 Glades Road, Boca Raton, FL 33431, USA}
\affiliation{National Astronomical Research Institute of Thailand, \\
260 Moo 4, T. Don Kaew, A. Mae Rim, Chiang Mai 50180, Thailand}

\author[0000-0002-2304-0908]{Vicki Sarajedini}
\affiliation{Department of Physics, Florida Atlantic University, \\
777 Glades Road, Boca Raton, FL 33431, USA}



\begin{abstract}
We present the results from a study of $\sim$9,600 Broad-Line selected AGN with host galaxies detected from the Sloan Digital Sky Survey Data Release 17 (SDSS DR17).  We compute ensemble variability statistics based on the comparison of the original SDSS photometric data with spectrophotometric measurements obtained days to decades later in the Sloan g-, r-, and i-bands. Galaxy and AGN templates have been fitted to the SDSS spectra to isolate the AGN component from the host galaxy. The sources have absolute magnitudes in the range -24 $<$ M$_i<$ -18 and lie at redshifts less than z $\sim$ 0.9. A variability analysis reveals that the anti-correlation between luminosity and variability amplitude continues down to log(L$_{bol}$ [erg s$^{-1}$]) = 43.5, demonstrating that the relationship extends by 4 orders of magnitude in AGN luminosity. To further explore the connection between AGN luminosity and variability, we determine the black hole mass and the accretion rate through measurement of the H$\beta$ line width and the monochromatic luminosity at rest-frame 5,100 $\AA$. Our results suggest that the accretion rate is the dominant parameter impacting the amplitude of variability and that the anti-correlation between accretion rate and amplitude extends to rates as low as 1\% Eddington. Moreover, we also identify an anti-correlation between variability amplitude and black hole mass, with the correlation appearing strongest among the AGN with low accretion rates.
\end{abstract}



\section{Introduction} \label{sec:intro}
Variability is a common feature of AGN and has been detected over a range of wavelengths and timescales \citep{matthews1963, sesar2006}. Since optical continuum radiation originates from the accretion disk, variability in the UV/optical regime is also thought to come from processes intrinsic to the accretion disk, such as inhomogeneities in the temperature at various radii \citep[e.g.,][]{ruan2014} and variations in the accretion rate \citep{pereya2006, zuo2012, sartori2019}. Reverberation mapping \citep[e.g.,][]{peterson2004} strongly implies that the accretion disk is the origin of the variability based on the lagging response of emission lines after continuum fluctuations. 

To probe the origin of the variability, correlations between the variability properties and the characteristics of the AGN have been explored. For example, the amplitude of variability has been found to correlate with several AGN parameters such as luminosity, redshift, wavelength, and possibly black hole mass \citep{hook1994, vandenberk2004, kelly2009, li2018, rumbaugh2018, suberlak2021, yi2022}. A particularly strong anti-correlation has been identified between variability amplitude and AGN luminosity, indicating that more luminous AGN vary less than fainter ones \citep[e.g.,][]{wilhite2008, zuo2012, simm2016}. 

Since AGN luminosity, accretion rate, and black hole mass are interdependent, considerable care has been taken to determine the driving factor for this anti-correlation. In a study of more than 7,500 quasars from the SDSS, \citet{zuo2012} found that the anti-correlation with accretion rate was more significant than that with luminosity when keeping other characteristics constant. More recently, \defcitealias{li2018}{L18}\citet[hereafter L18]{li2018} and \citet{yu2022} found that the amplitude of variability, particularly the long-term amplitude, is primarily determined by the accretion rate. Likewise, \citet{rumbaugh2018} found that extreme variability AGN (those with fluctuations greater than a magnitude) generally have lower accretion rates than normal quasars with similar luminosities and suggest that the variability mechanism at very low accretion rates may be different than that for typical quasars. 

To explore the relationship between various AGN parameters at fainter luminosities than previously studied,\defcitealias{gs2014}{GS14} \citet[hereafter GS14]{gs2014} selected a sample of $\sim$5,000 galaxies from the SDSS DR7 having broad emission lines and extended morphologies. By requiring that the sources have galaxy-like morphologies, the sample was not biased against intrinsically faint AGN with significant contributions from host galaxies. Using the approach of \defcitealias{vandenberk2006}{VB06}\citet[hereafter VB06]{vandenberk2006}, they separated the AGN and host galaxy light using linear combinations of eigenspectra templates for quasars and galaxies. The variability was quantified as the difference between magnitudes measured from the source photometry and follow-up spectroscopy in the Sloan g-, r-, and i-bands. They constructed an ensemble Structure Function (SF) for these sources and confirmed the anti-correlation found among quasars for AGN extending $\sim$2 magnitudes fainter than previously studied. While the amplitude of the variability was found to increase to the faintest luminosities, the slope of the SF became slightly shallower than that measured for more luminous QSOs.

In this paper, we expand on the study of \citetalias{gs2014} using the SDSS DR17, the data release including observations through January 2021. We identified close to 9,600 galaxies with broad emission lines, extended morphologies, and sufficient spectroscopic and photometric signal-to-noise (S/N) necessary to model the host galaxy contribution and measure various AGN properties including the ensemble SF. We use this large sample of AGN to explore correlations between the ensemble variability properties and AGN characteristics such as accretion rate and black hole mass at lower luminosities than previously studied. 

In Section \ref{sec:dataset}, we describe the selection of the data set utilized for this study as well as the control set of normal galaxies to characterize the photometric noise. We explain the principle component analysis (PCA) decomposition used to determine the host galaxy fraction for each AGN and the procedure used to correct the variability amplitude for the host galaxy contribution. We also discuss the sample trimming based on the broad emission line measurements, spectroscopic signal-to-noise, and host galaxy fraction. In Section \ref{sec:analysis}, we measure the broad emission lines for the AGN component of each source and determine the black hole masses and accretion rates for our sample. In Section \ref{sec:ensemble variability}, we compute the ensemble Structure Function for low-luminosity AGN and perform multidimensional fits of the AGN parameters and variability amplitudes. We discuss the implications of these correlations in the low-accretion rate, low-luminosity AGN regime for AGN in Section \ref{sec:discussion} and present conclusions in Section \ref{sec:conclusions}. Throughout the paper, we use photometric data from SDSS in their system \citep{Lupton1999}, which is similar to the AB photometric system, and $\Lambda$CDM cosmology with $\Omega_M$ = 0.30, $\Omega_\Lambda$ = 0.70, and $H_0$ = 70 $km \: s^{-1} \: Mpc^{-1}$.


\section{Data Set} \label{sec:dataset}

\subsection{The Survey}\label{sec:sdss17}

The Sloan Digital Sky Survey is a photometric and spectroscopic survey covering almost 15,000 square degrees or about one-third of the entire celestial sphere in five broad bands, namely u, g, r, i, and z. We utilize data from the Data Release 17 (DR17), which includes imaging and spectroscopy data obtained over 19 years up to and including January 2021 \citep{sdss-dr17}. This is the final data release for SDSS-IV which contains more than a billion catalog objects, including more than 200 million galaxies. Among those, there are almost 3 million galaxies and 1 million quasars with spectroscopy. The photometric depth (5$\sigma$) of the survey is 22.15, 23.13, 22.70, 22.20, and 20.71 magnitudes for u-, g-, r-, i-, and z-bands, respectively. 

\subsection{Selection Criteria for SDSS DR17} \label{sec:selection} 
   
We follow a similar approach as \citetalias{gs2014} and perform the CasJobs' SQL query from the SDSS DR17 to gather all relevant parameters such as object ID, coordinates, PSF-corrected flux and magnitude, fiber flux and magnitude, spectroscopic flux and magnitude, median signal-to-noise in the respective band, observed Modified Julian Date (MJD), absolute magnitude, extinction, redshift, and K-correction factor. We select sources that are defined as morphologically extended (photometric type = 3) and display at least one broad emission line in the spectrum (spectral class = ``QSO''), aiming toward the lower-luminosity AGN. We further required that the broad line measure at least 1,000 km $s^{-1}$ and no more than 10,000 km s$^{-1}$. This ensures that the spectral classification is based on the presence of at least one robustly-detected broad emission line and that the emission line width is consistent with Doppler broadening of gas around black holes up to $\sim$10$^9$ M$_{\odot}$.  

We impose a redshift constraint of z$\leq$0.84 to ensure that the spectra contain the entire broad H$\beta$ emission feature. The H$\beta$ emission line is useful for our determination of supermassive black hole mass (M$_{SMBH}$). The redshift cut guarantees sufficient overlaps between the observed spectra and the templates for high-quality fitted results. The total number of AGN that pass the above criteria is 32,828 sources. 


We also require a sample of ``normal" non-varying galaxies as a control group to gauge the photometric offsets and noise between different epochs. We select 2,450,896 normal galaxies (extended sources with no broad emission lines) from the entire SDSS DR17 with the same set of criteria except setting the spectral class to ``GALAXY" to exclude objects with broad emission lines. We also apply the same redshift cut to this control group.
 
\subsection{Measuring Variability} \label{sec:variability}

We calculate the variability of our sample using the differences in photometric magnitudes between two epochs; the imaging data and the spectroscopic data obtained for each galaxy in SDSS. Thus the baseline between two epochs can span from days to a decade. The spectroscopic reduction pipeline \citep{stoughton2002} calculates synthetic spectrophotometry for the g-, r-, and i-filters. The magnitude difference is computed as $\Delta$g = g$_{sp}$ - g$_{ph}$ and similarly for other bands. Note that this is in the opposite direction of the magnitude differences from \citetalias{gs2014}, thus $\Delta$g = -$\Delta$g$^{GS14}$. We drop the u- and z-bands from further analysis due to the lack of overlaps between the templates and the observed spectra in both bands for higher redshift bins of the sample. 

During the summer of 2009, there was a change in the multi-object, fiber-fed spectrograph of SDSS \citep{smee2013}. The spectrograph was upgraded for the Baryonic Oscillation Spectroscopic Survey as a part of the SDSS-III \citep{dawson2013}. The change of spectrograph, however, resulted in significant differences between the spectrophotometry with the new instrument and the imaging photometry obtained from the previous version of the instrument. For this reason, we compute systematic offsets and noise characteristics for our sources in two separate groups, namely Group I and Group II. Group I contains AGN and galaxies with spectra taken before the Modified Julian Date (MJD) of 55,000 days (i.e., before the upgrade). Group II contains those with spectra taken after the upgrade. There are 13,112 and 19,716 AGN in Group I and Group II, respectively. Due to the different observing strategies between SDSS-I through -II and SDSS-III through -IV, Group II sources are at higher redshifts and, thus, fainter and lower S/N values. 

To properly determine the magnitude difference between the two epochs, we use the SDSS aperture photometry within 3-arcsecond and 2-arcsecond diameters to closely match the fiber sizes of the spectroscopic observations of Group I and Group II samples, respectively. Since the spectrophotometry was calibrated using PSF-fitted magnitudes \citep{adelman2008}, which includes light beyond the aperture of the spectroscopic fiber, the photometric offset varying significantly with S/N was observed among both Group I and Group II's AGN and normal galaxies. 

\begin{figure}[h] 
\centering
\includegraphics[width=0.95\linewidth]{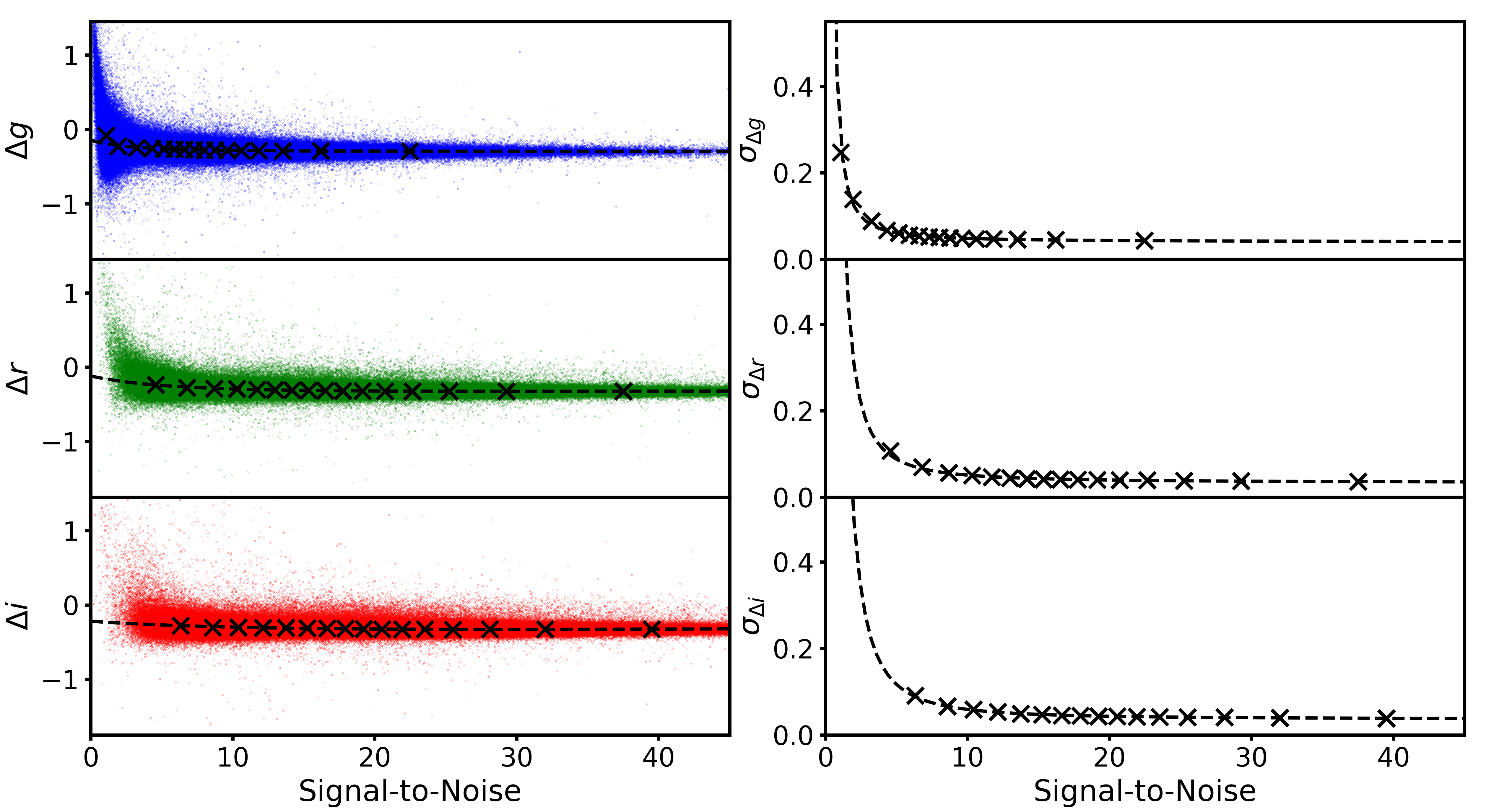}
\caption{{\bf Left:} Change in magnitude differences as a function of spectroscopic S/N of Group I normal galaxies sample (control sample) in g-, r-, and i-bands, respectively. {\bf Right:} The corresponding standard deviations (sigma values) of the corrected magnitude differences in each band. The black crosses represent the centroids of magnitude differences and the standard deviations in each bin for all bands. While the black-dashed lines are the best-fitted Laurent polynomials.}\label{figure1}
\end{figure}

\begin{figure}[h]
\centering
\includegraphics[width=0.95\linewidth]{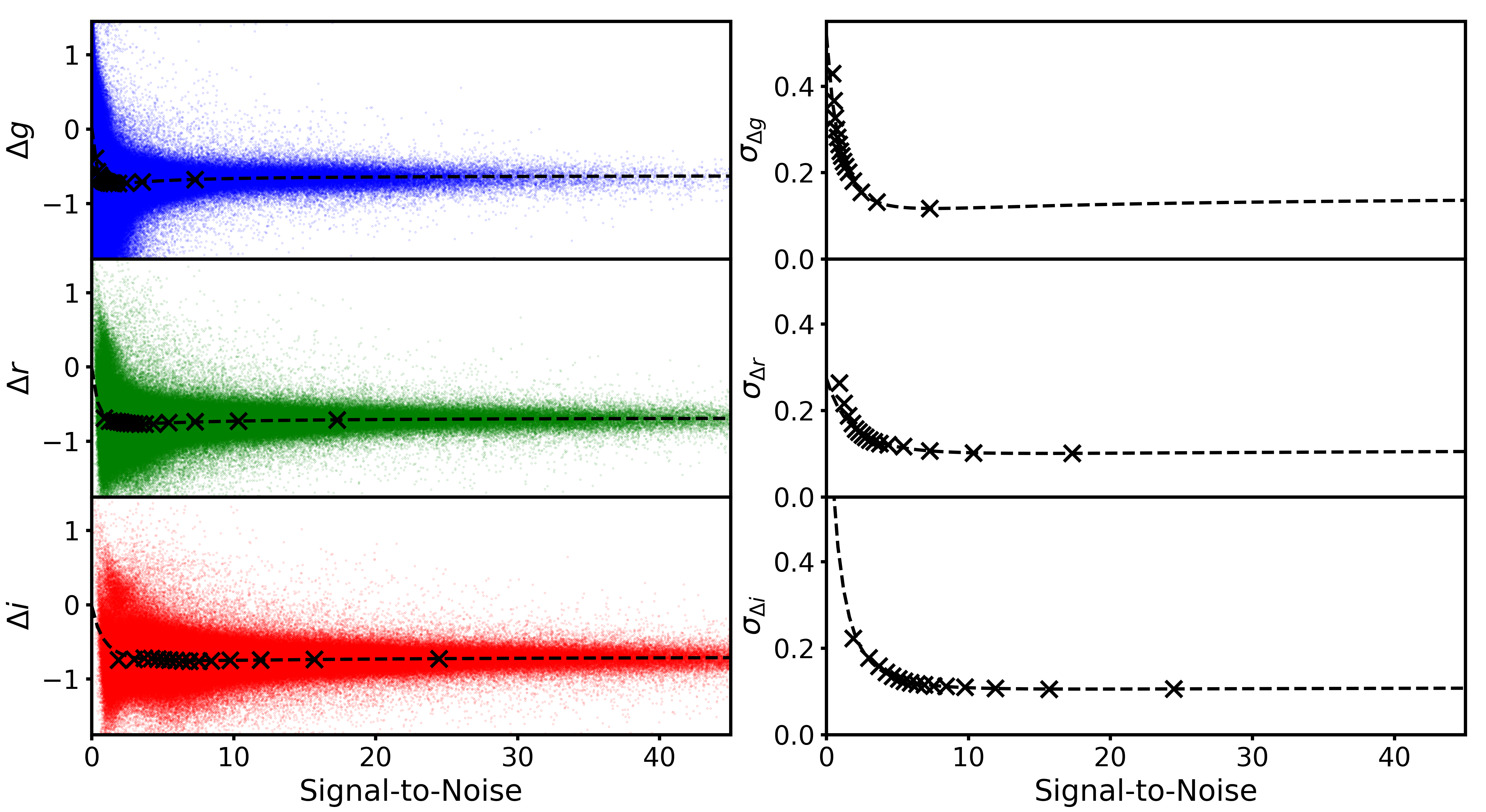}
\caption{{\bf Left:} Change in magnitude differences as a function of spectroscopic S/N of Group II normal galaxies sample (control sample) in g-, r-, and i-bands, respectively. {\bf Right:} The corresponding standard deviations (sigma values) of the corrected magnitude differences in each band. The black crosses represent the centroids of magnitude differences and the standard deviations in each bin for all bands. While the black-dashed lines are the best-fitted Laurent polynomials.}\label{figure2}
\end{figure}

To quantify the offset, we divide the sources into 16 S/N bins spread across the middle 95 percent of the sources (corresponding to the 2$\sigma$ confidential interval) with an equal number of sources per bin. In each filter of Group I and II, the centroids and standard deviations of the magnitude difference distribution are measured. The left panels of Figure \ref{figure1} and \ref{figure2} show the distributions of magnitude differences as a function of S/N for Group I and Group II of the control sample (normal galaxies), respectively. The black crosses represent the median values of the magnitude differences in each bin. In the high S/N regime (S/N$>$5), the offset is -0.32 and -0.70 for Group I and Group II, respectively. 

We fit a second-order Laurent polynomial to these values to determine the offset as a function of S/N for each group in different filters. The offset is then applied to the magnitude difference for each normal galaxy to center the magnitude difference distribution at zero. After centering in each band, we measure the spread using the standard deviation as a function of S/N for the control sample. This represents the ``photometric noise'' expected for non-varying galaxies in our survey. The right panels of Figure \ref{figure1} and \ref{figure2} show the Gaussian fits to the distributions of magnitude differences as functions of S/N. The photometric noise decreases almost exponentially as the S/N values rise. The Group II control sample has an overall higher photometric noise due to the change in spectrograph and observing strategies mentioned earlier. We also fit a second-order Laurent polynomial to these data points in both groups and use the derived functions in Section \ref{sec:ensemble variability} to obtain the characteristic photometric noise, $\sigma_{phot}$, for each AGN in the sample.  

\begin{figure}[h] 
\centering
\includegraphics[width=0.95\linewidth]{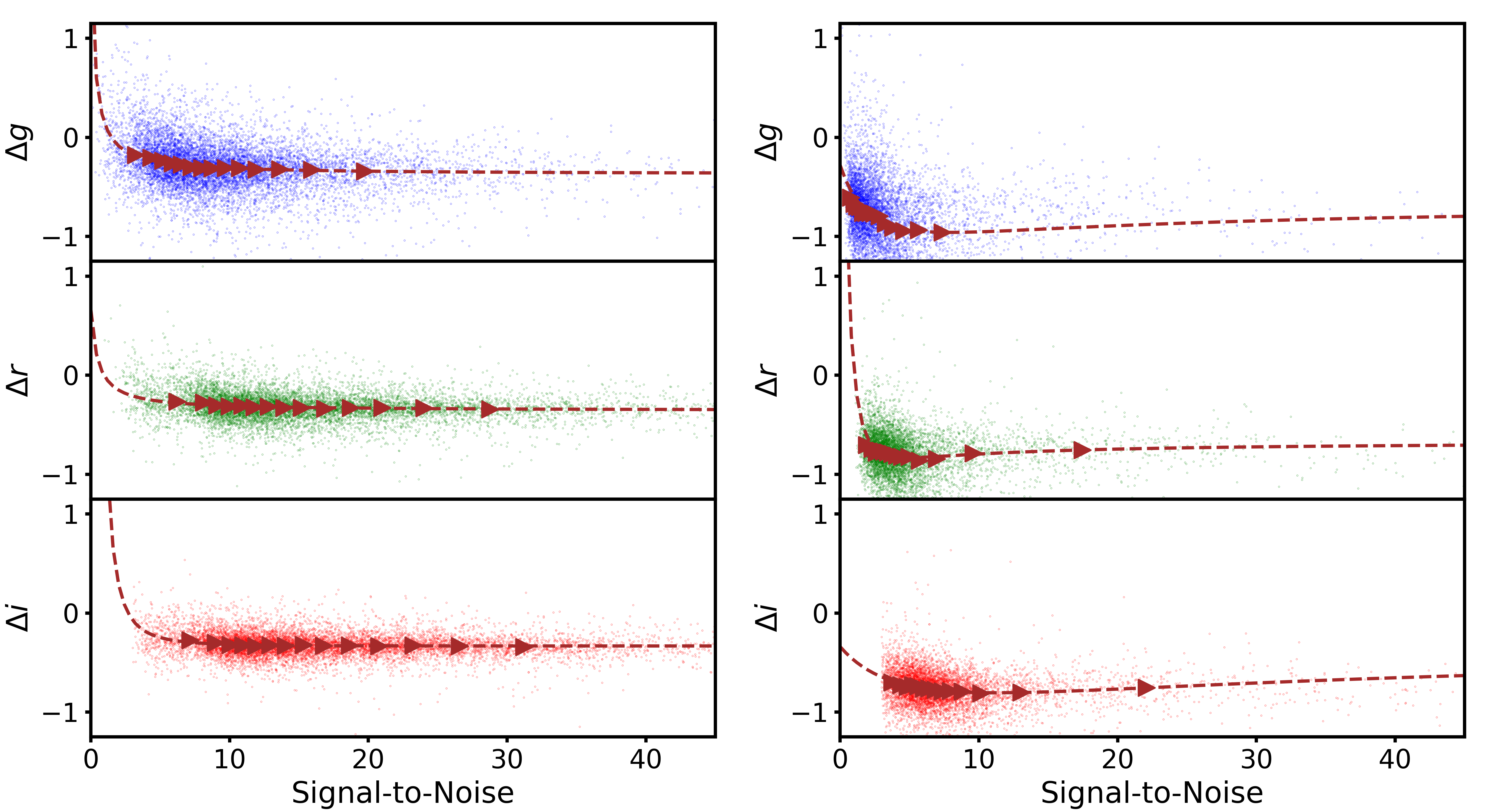}
\caption{{\bf Left:} Change in magnitude differences as a function of spectroscopic S/N of Group I AGN sample in g-, r-, and i-bands, respectively. {\bf Right:} Change in magnitude differences as a function of spectroscopic S/N of Group II AGN sample in g-, r-, and i-bands, respectively. The brown triangles represent the centroids of magnitude differences in each bin for all bands for both Group I and Group II AGN. While the brown-dashed lines are the best-fitted Laurent polynomials.}\label{figure3}
\end{figure}

The offsets of magnitude differences are found to be slightly higher for the AGN sample compared to the normal galaxies sample in all bands. The same characteristics were also noted by \citetalias{gs2014} and may be due to differences in the morphologies of AGN (e.g., a higher central concentration of light) compared to normal galaxies. Therefore, we compute the magnitude offset separately for the AGN sample in both groups and divide the samples into S/N bins following the procedure used for the normal galaxies. We determine the centroids of the magnitude differences shown as brown triangles in Figure \ref{figure3}. The left and right panels of Figure \ref{figure3} show the magnitude differences offsets of AGN in Group I and Group II, respectively. For instance, at the S/N of 10, the offsets are -0.31 and -0.95 for AGN in Group I and Group II, respectively. We, again, fit a second-order Laurent polynomial to determine the offset as a function of S/N for each group in each filter. This offset is applied to the magnitude differences for all AGN in our sample to center its distribution at zero. The variability analysis in Section \ref{sec:ensemble variability} makes use of these offset-corrected magnitude differences.

\begin{figure}[h] 
\centering
\includegraphics[width=0.95\linewidth]{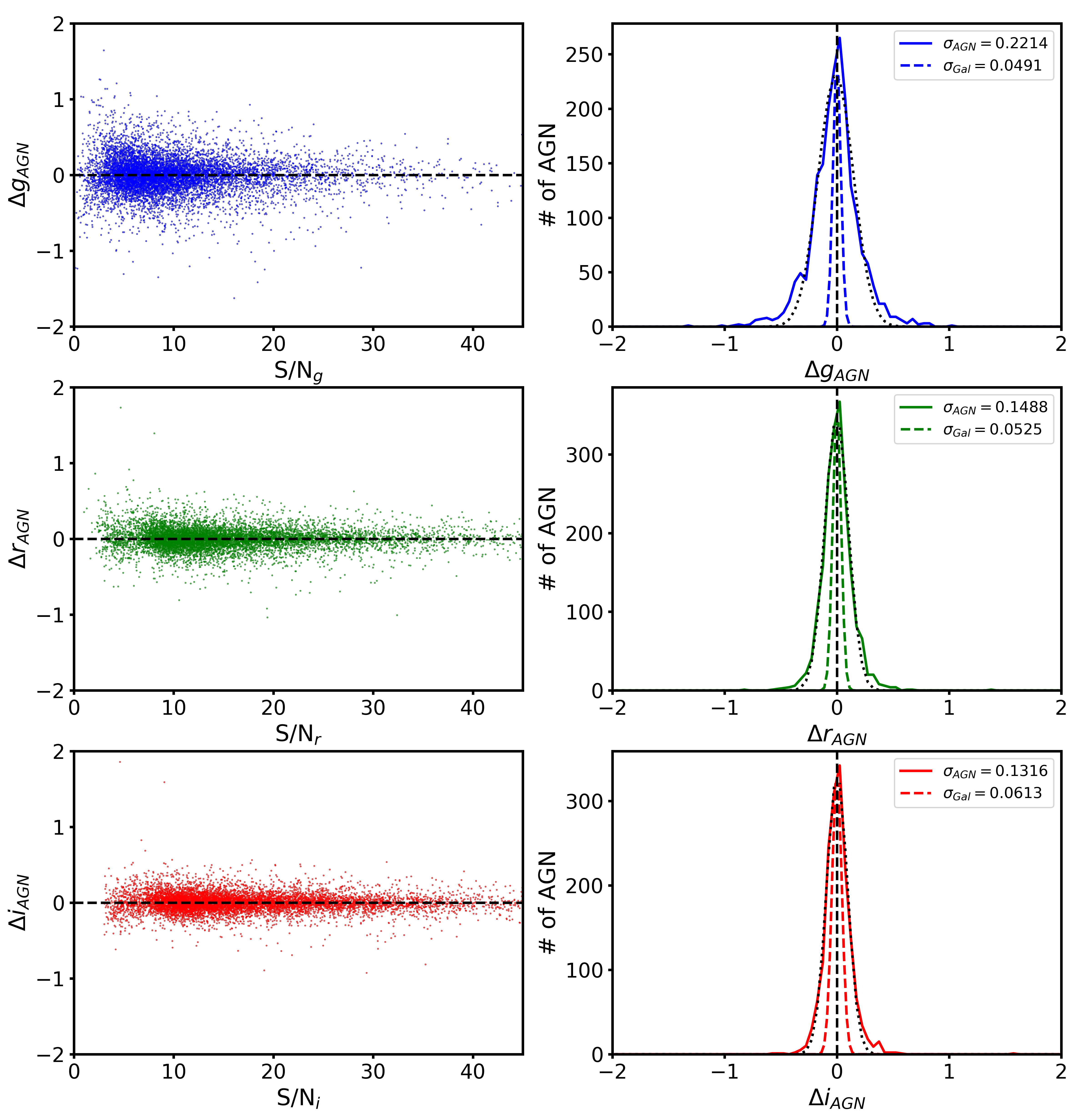}
\caption{{\bf Left:} Group I AGN sample's magnitude differences corrected for the systematic offset. {\bf Right:} The distribution of magnitude differences at S/N of 10 for the AGN sample (solid line) and fit (dotted line) shown with the control sample (dashed line)}\label{figure4}
\end{figure}

\begin{figure}[h] 
\centering
\includegraphics[width=0.95\linewidth]{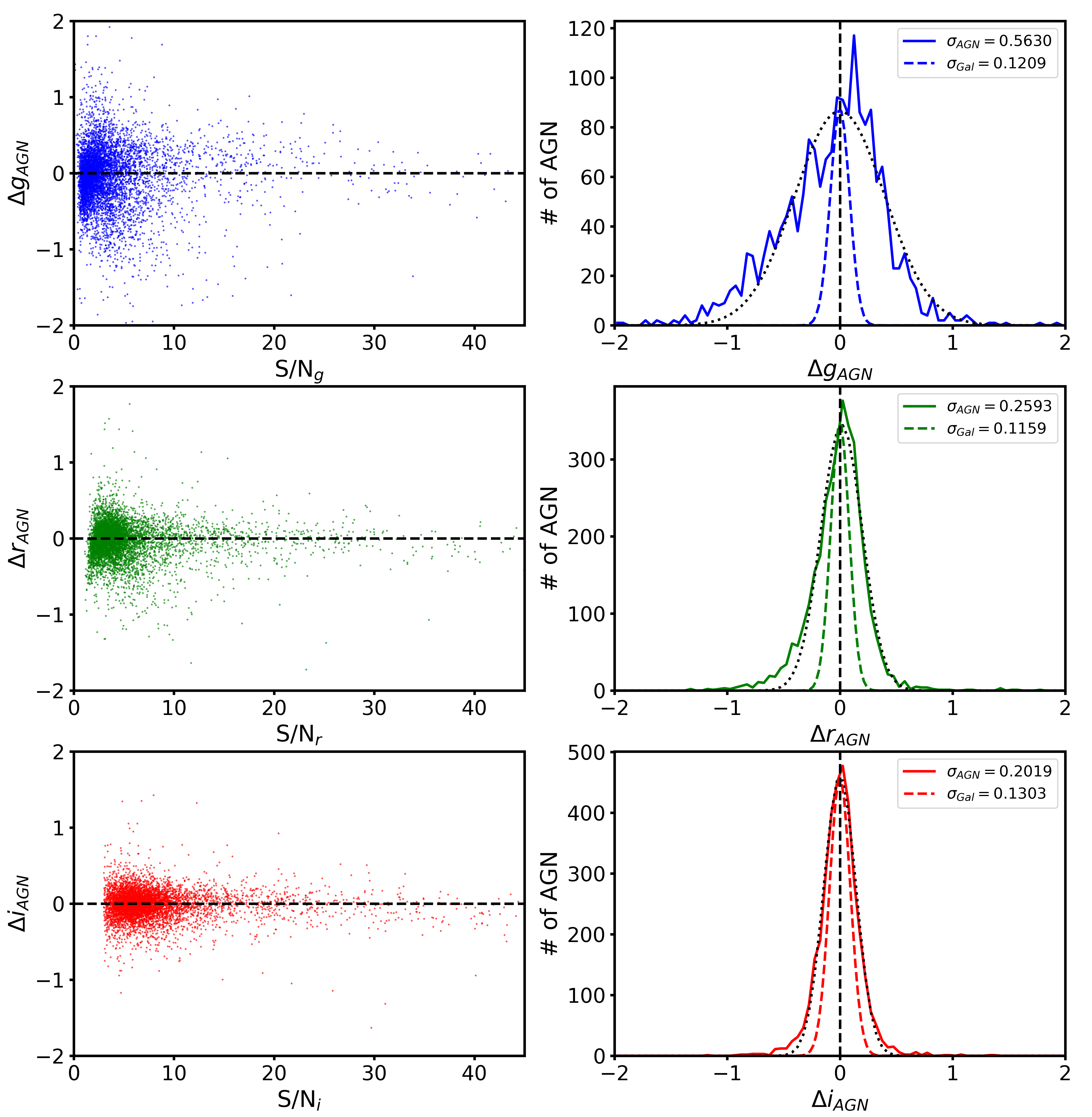}
\caption{{\bf Left:} Group II AGN sample's magnitude differences corrected for the systematic offset. {\bf Right:} The distribution of magnitude differences at S/N of 5 for the AGN sample (solid line) and fit (dotted line) shown with the control sample (dashed line)}\label{figure5}
\end{figure}

To demonstrate the intrinsic variability among the AGN compared to the non-varying galaxies, Figures \ref{figure4} and \ref{figure5} show the offset-corrected magnitude differences for the AGN sample (left panels) and the distribution of these values around the key S/N levels of 10 and 5 (right panels) with the distribution of the normal galaxy sample at the same S/N shown with a dashed line for Group I and Group II objects, respectively. A fit to the distribution of magnitude differences (dotted line) yields a Gaussian 1$\sigma$ value of 0.22 for the AGN sample and 0.05 for the normal galaxies in the g-band for Group I. Similarly, the fits are 0.15 and 0.05 for the r-band and 0.13 and 0.06 for the i-band. In Group II, since the overall targets are fainter, the bulk of S/N distribution shifted toward lower S/N values. Thus, we show the distribution of the magnitude differences in Group II at a S/N of 5. The significantly larger magnitude differences for the AGN sample in all bands and both groups demonstrate their variable nature beyond the photometric noise. 



\subsection{PCA Decomposition} \label{sec:pca}

To properly analyze the variability properties of our sample, we must first take into consideration the contamination of light from the host galaxy. Since we have selected the morphologically extended galaxies, the flux contribution from the host galaxy is non-negligible. Assuming that light from the underlying galaxy is not varying, the host galaxy's flux will dilute the observed variability of the AGN. The effect is more severe as the host galaxy contribution rises. To gauge the contribution from the host galaxy, we needed to separate the observed spectrum into two components consisting of the AGN and the host galaxy. We follow the formalism in deconstructing the AGN and host galaxy components as described in \citetalias{gs2014} and \citetalias{vandenberk2006}. Any source spectrum can be expressed as a linear combination of eigenspectra as shown here:

\begin{equation}
f^R_\lambda = \sum_{k=1}^N a_ke_k(\lambda)
\label{eq1}
\end{equation}

\noindent where f$^R_\lambda$ is the reconstructed flux density as a function of the wavelength of the interested component, while a$_k$ and e$_k(\lambda)$ are the k-order eigencoefficient and the corresponding eigenspectra, respectively. 

We adopt the Principal Component Analysis or PCA fitting and decomposition routine from \citet{Hao2005}, previously developed to analyze SDSS spectra for quasars and galaxies. However, we have modified the routine for use in Python utilizing SciPy's optimization with weights and penalties to fit all eigenspectra to the observed spectrum simultaneously. The eigenspectra for galaxies and quasars are characterized and made available by \citet{yip2004a, yip2004b}. For the eigenspectra of quasars, we pick the group that represents the brightest population at the lowest redshift range. We choose the low-redshift bin, namely ``ZBIN 1'', which covers the redshift range between $0.08<z<0.53$, overlapping with the majority of AGN in our sample. We also select the highest luminosity bin, namely ``C1'', with the absolute magnitude of quasars between -26 $\leq$ M$_i \:\leq$ -24 \citep{yip2004b}. The latter selection guarantees that the quasar eigenspectra will have minimal contamination from the host galaxy. Our redshift limit of z$\leq$ 0.84 ensures sufficient overlap between the AGN's rest-frame spectra and the eigenspectra for a high-quality fit and decomposition result.


The number of eigenspectra or PCA bases needed to fit and decompose the observed AGN spectra efficiently can be varied from a few to fifty vectors. However, many studies \citep[e.g.,][]{connolly1995, connolly1999, reichardt2001, yip2004b, gs2014, davies2018, fagioli2020} have shown that the majority of the information in the spectra is contained within the first few vectors. For our purpose of gauging the contributions of AGN and host galaxy, we determine that using 5 galaxy eigenspectra and 10 quasar eigenspectra to perform the PCA decomposition is sufficient (e.g.,\citetalias{gs2014}; \citetalias{vandenberk2006}). To separate the components, we assign weights (w($\lambda$) = 1/$\sigma$($\lambda$)$^2$) to the observed flux density; where $\sigma$($\lambda$)$^2$ is the variance of each flux density value. The w($\lambda$) is of the order unity except for around prominent emission features of the AGN \citep{stoughton2002}. The regions within $\pm$ 5,000 km s$^{-1}$ of H$\alpha$, H$\beta$, H$\gamma$, and H$\delta$ emission lines, and $\pm$ 600 km s$^{-1}$ of [OII $\lambda\lambda$ 3727,3729] and [OIII $\lambda\lambda$ 4959,5007] are given weights of 0.001 to penalize the peaky emission features with high S/N \citep{Hao2005}. This approach mitigates the issue of overestimating the continuum levels in favor of fitting strong emission features. Thus, we ensure that the overall contributions of each component are as accurate as possible. 

\begin{figure}[h] 
\centering
\includegraphics[width=0.95\linewidth]{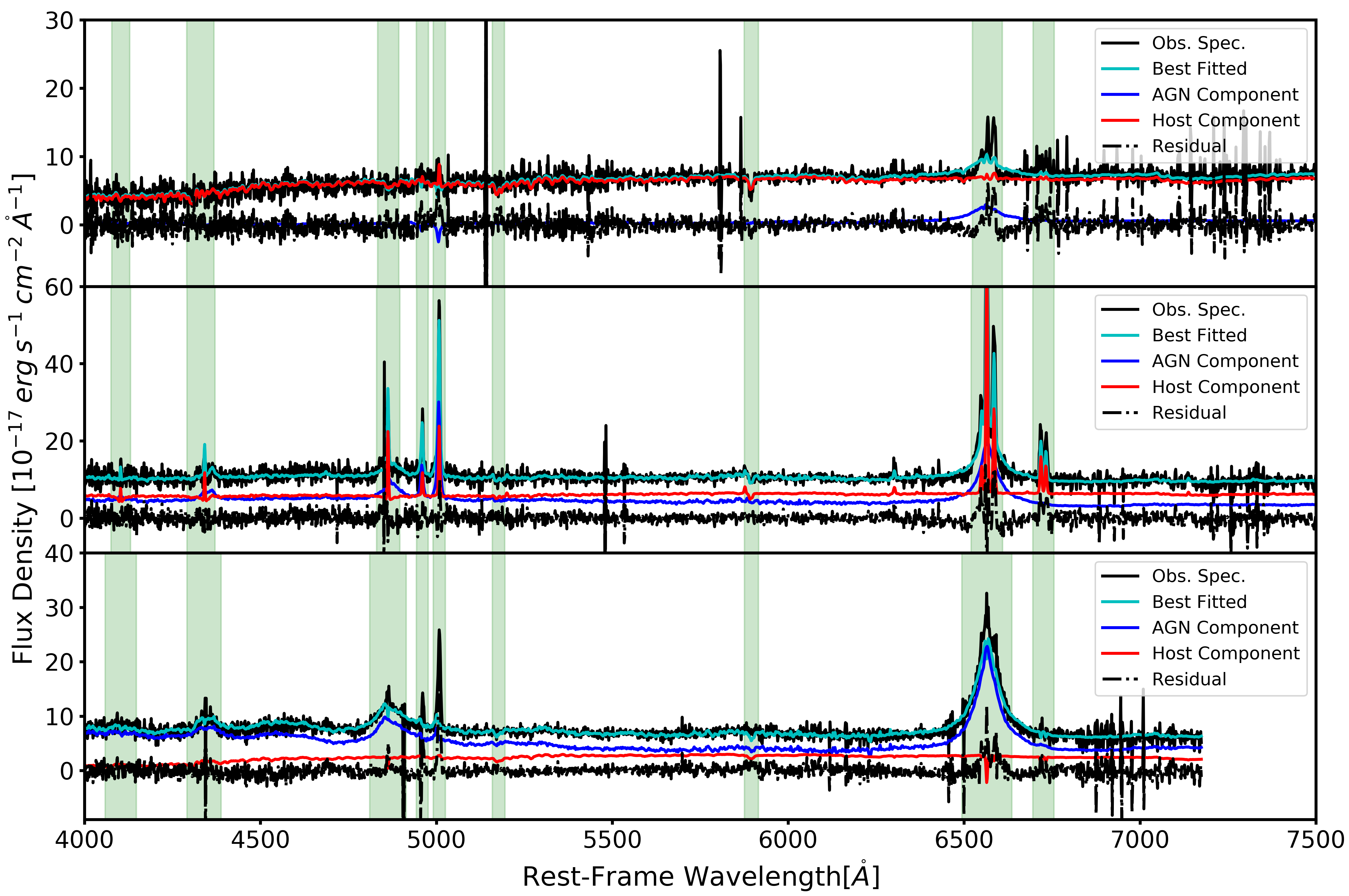}
\caption{From top to bottom, the panels show typical PCA decomposition results when performed on host-dominated, 50\%, and AGN-dominated spectra, respectively. The black solid lines are the observed AGN spectra. The cyan, blue, and red solid lines are the best-fitted total, AGN, and host galaxy components from the PCA decomposition, respectively. The black-dashed lines are the residuals of the fits. The green-shaded regions represent the areas on the spectra that have been masked out before calculating host fraction values of AGN. 
}\label{figure6}
\end{figure}

We measure the contribution from the host galaxy in a similar fashion as \citetalias{vandenberk2006} and \citetalias{gs2014}. This is known as the ``Host Fraction'' or ``$\Psi$''. The $\Psi$ value of each object is the ratio between the integrated observed flux of the host galaxy component and the observed total flux (AGN + host) within the corresponding Sloan filter. Since the host fractions are derived from the reconstructed spectra, we can calculate $\Psi$ in both observed- and rest-frames for all Sloan filters. $\Psi$ can be expressed as follows:

\begin{equation}
\Psi_j = \frac{\int_{\lambda1}^{\lambda2} R^j_\lambda M^j_\lambda  f^{Host}_\lambda d\lambda}{\int_{\lambda1}^{\lambda2} R^j_\lambda M^j_\lambda (f^{Host}_\lambda + f^{AGN}_\lambda) d\lambda}
\label{eq2}
\end{equation}

\noindent where R$^j_\lambda$ represents the filter response as a function of wavelength in an arbitrary filter ``j'', and $\lambda$1 and $\lambda$2 are cut-on and cut-off wavelengths of the corresponding Sloan filter. While, f$^{AGN}_\lambda$ and f$^{Host}_\lambda$ are the reconstructed flux densities of the host galaxy and the AGN components, respectively. Finally, M$^j_\lambda$ is the mask on the specific wavelengths within the corresponding filter. The mask blocks out the region with the size of 2.5$\times$ full width at half maximum (FWHM) of the broad emission lines (or $\pm$3$\sigma$) of the AGN component and the region with the size of 1,000 km s$^{-1}$ of the narrow emission lines of the host galaxy component. Thus, the host fraction value calculated in each filter is essentially the ratio between the average continuum levels of the host galaxy and the combined AGN + host galaxy. A few examples of the PCA decomposition results are illustrated in Figure \ref{figure6}. From the top to bottom panels of Figure \ref{figure6}, we show three AGN having a median S/N in the r-band around 10 and the corresponding host galaxy fraction values in the observed frame, $\Psi_r$, of $>$0.8, $\sim 0.5$, and $<$0.2, representing sources where the host galaxy is more than 80\%, around 50\%, and less than 20\% of the total light. The masked areas are indicated by green-shaded regions.

With the host fractions determined, we can calculate the magnitude difference between a pair of epochs for the AGN component corrected for the dampening effect of the non-varying host galaxy. The magnitude difference between 2 epochs of a host-subtracted AGN in any arbitrary band, ``j'', can be expressed as follows:

\begin{equation}
\begin{split}
|\Delta m_j^{AGN}| = 2.5|log(\frac{f^{spec^{AGN}}_j}{f^{phot^{AGN}}_j})| \\
= 2.5|log[\frac{(1-\Psi_j)(10^{k_j} - \Psi_j \phi_j) }{(1-\Psi_j \phi_j)(10^{-k_j} - \Psi_j)}]|
\end{split}
\label{eq3}
\end{equation}

\noindent where $\phi_j$ = 10$^{\Delta m_j}/{-2.5}$ is the corrected spectrophotometry to photometry flux ratio (f$_j^{spec}$/f$_j^{phot}$) and k$_j$ = -0.2$\times$($\Delta$m$_j$). The value $\Delta$m$_j$ is the offset-corrected total apparent magnitude difference (m$^{spec}$ - m$^{phot}$). The values of f$_j^{spec}$ and f$_j^{phot}$ are fluxes in any arbitrary band, j, from spectrophotometric and photometric epochs, respectively. Note that we cannot use the same host fraction values for both epochs since it contradicts the initial assumption that the host galaxy component is non-varying. Thus, we allow the $\Psi$ value to vary (in the opposite direction) as the brightness of the AGN varies, or $\Psi'$ = $\Psi\times$f$_j^{spec}$/f$_j^{phot}$. Because the corrected magnitude differences of the AGN components are quite sensitive to the values of host fractions, we need to be certain that the estimated host fraction values obtained through this procedure are accurate and reliable. We verify the reliability of the host fraction (and ultimately AGN fraction) determination in the following section. 

\subsection{Reliability of AGN/host galaxy fraction from PCA fitting}\label{sec:reliability}

\begin{figure}[h] 
\centering
\includegraphics[width=0.95\linewidth]{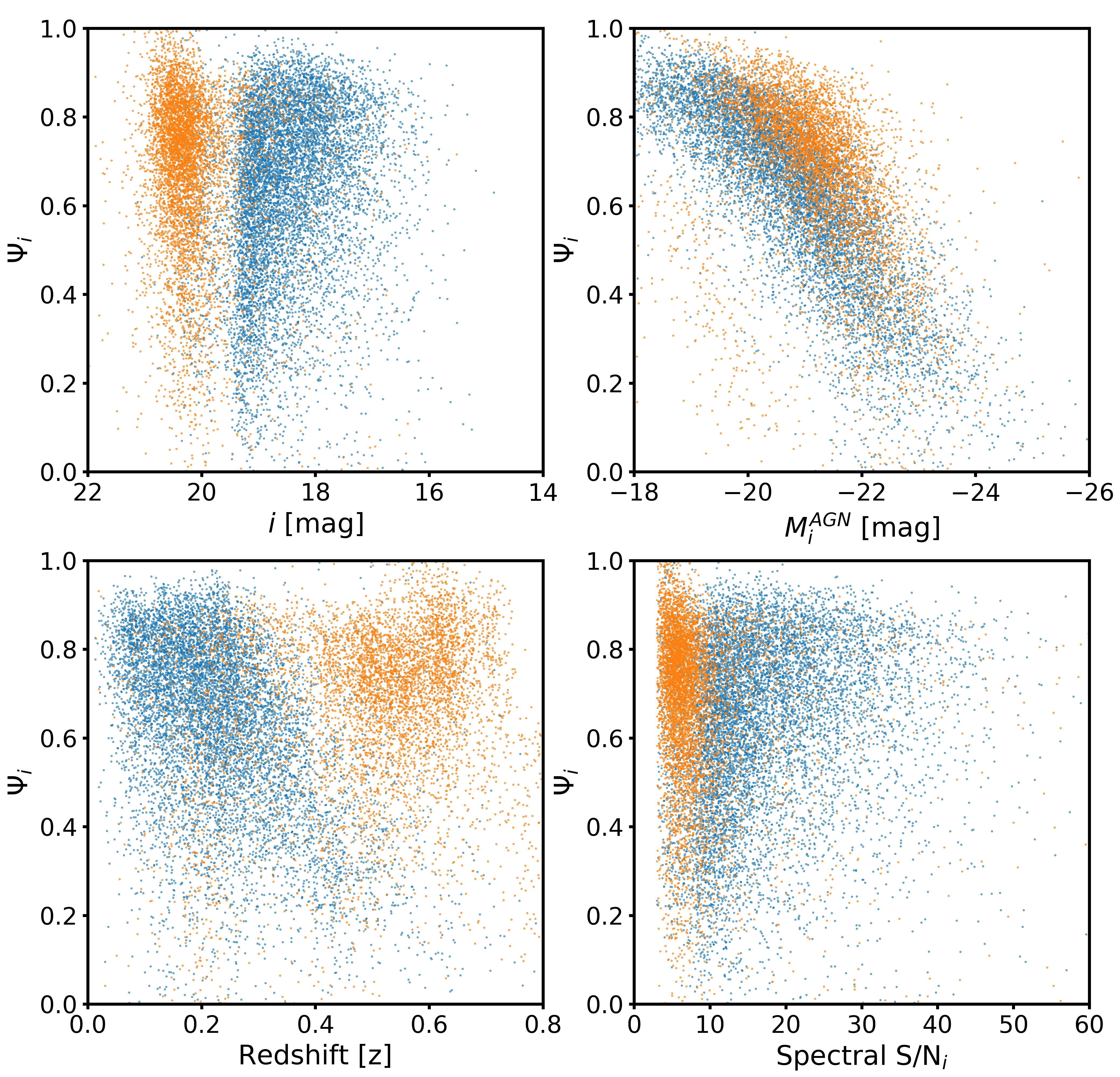}
\caption{The distributions of host galaxy fraction values in i-band, $\Psi_i$, with respect to apparent i-band magnitude ({\bf top left}), absolute magnitude in i-band of AGN component ({\bf top right}), redshift ({\bf bottom left}), and median spectroscopic S/N in i-band ({\bf bottom right}). The blue and orange circles represent the AGN with spectroscopy taken before and after the upgrade of the SDSS spectrograph (i.e., Group I and II AGN), respectively.
}\label{figure7}
\end{figure}

Figure \ref{figure7} shows the host galaxy fraction in the i-band, $\Psi_i$, as a function of various observational parameters. Here we plot $\Psi_i$ versus the apparent magnitude from the SDSS photometry, the extinction corrected absolute magnitude of the AGN component as determined from the PCA fit, the source redshift, and the SDSS spectroscopic S/N in the i-band. We show the AGN from Group I and Group II as blue and orange points, respectively. The upper and lower left panels highlight the post-2009 survey differences, targeting higher redshift and fainter sources generally. The upper right panel shows that the luminosity of the AGN component is fainter where the host galaxy dominates, extending to M$_i\sim$-18. There are no significant trends with $\Psi_i$ and S/N in the i-band, indicating that the full range of AGN-to-host galaxy ratios can be detected at all S/N values. 

\begin{figure}[h] 
\centering
\includegraphics[width=0.95\linewidth]{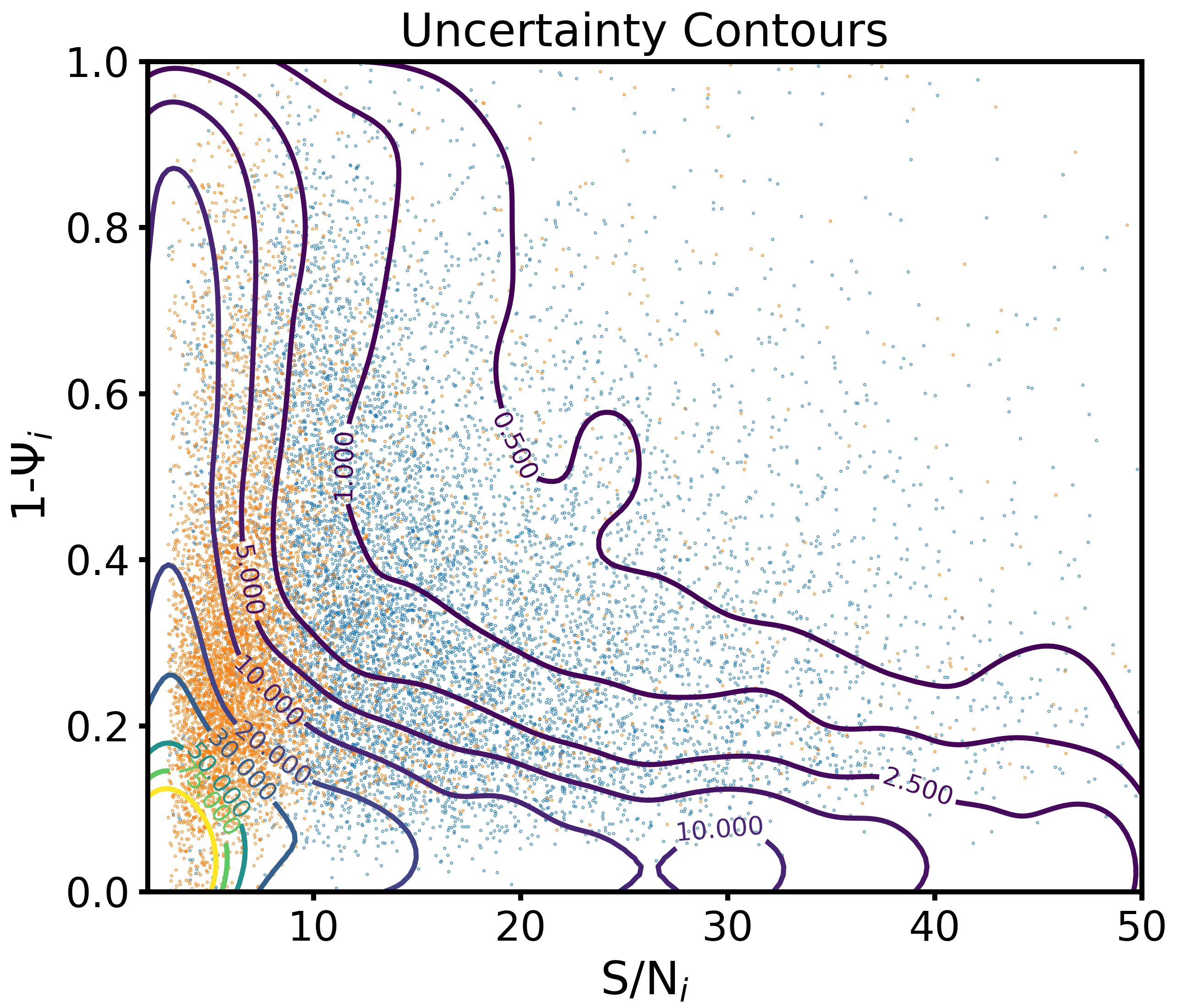}
\caption{The reliability contours exhibiting uncertainty levels in recovering AGN fraction in i-band ($1-\Psi_i$) in response to the initial $1-\Psi_i$ and $S/N_i$ values. The label values indicate the average uncertainty in the fitted AGN fraction in percent. The blue and orange circles represent the AGN with spectroscopy taken before and after the upgrade of the SDSS spectrograph (i.e., Group I and II AGN), respectively.
}
\label{figure8}
\end{figure}

Since the $\Psi$ parameter has a significant influence on the determination of the AGN magnitude and variability correction, we perform additional simulations to ensure that all sources included in our study have reliable estimates for host fraction. To test the accuracy of our PCA decomposition routines, we use the quasar and galaxy eigenspectra to generate a synthetic spectrum of an AGN. We tune the eigenvalues of the eigenspectra to match the range of $\Psi_i$ values from 0.0 to 1.0 and impose fluctuations of the flux densities for the synthetic spectra with a range of S/N$_i$ values from 2 to 50, consistent with the range of our sources. We simulate spectra in 40 bins of $\Psi_i$ and 32 bins in S/N$_i$, equivalent to 1,280 cells on the $\Psi_i$-S/N$_i$ space. In each cell, there are 25 synthetic spectra with random redshift values, drawn from the same redshift distribution of our sources, with the $\Psi_i$ and S/N$_i$ values drawn from a uniform distribution bounded by the cell's edges. Thus, we generate a total of 32,000 simulated AGN spectra. The simulated spectra are fed through our PCA decomposition routines. The deviations from the input values of $\Psi_i$ are recorded and averaged for each cell. Next, we increase the sampling along the $\Psi_i$ and S/N$_i$ grids to 100 and 192 bins (or 2.5$\times$ and 6$\times$), respectively. The up-sampling is done by 2D interpolation with a linear estimator. Finally, we smooth the grids with a 2D Gaussian Kernel using a standard deviation of 4 (up-sampled) cells. The resulting contours of the average uncertainty of the recovered AGN fraction in the i-band as a function of the input AGN fraction, or (1-$\Psi_i$), and S/N$_i$ is shown in Figure \ref{figure8}. This figure is similar to the lower right panel of Figure \ref{figure7}, but with the y-axis flipped so that higher AGN fractions are at the top and host galaxy-dominated sources are at the bottom of the figure.

As one expects, the simulation shows that the least reliably recovered AGN fraction values have low S/N and low AGN fractions. Based on this result, we choose to include only the AGN with reliability greater than 95\% in recovering the correct AGN fraction. This eliminates sources to the left and below the line labeled ``5.000'', removing most sources with S/N$_i<$7 and having AGN fractions less than 20\% (or host galaxy fraction larger than 80\%). Due to these cuts, a substantial number of Group II AGN are removed from our sample. Based on the results of this simulation, we continue the variability analysis with 9,798 AGN (7,649 sources in Group I and 2,149 sources in Group II). 

\section{Analysis}\label{sec:analysis}

\subsection{Broad Emission Lines Extraction} \label{sec:emissionline} 

With our sample of reliably decomposed spectra, we carry out the remaining analysis using the host galaxy-subtracted spectrum for each AGN. Since we are interested in determining the black hole masses, we carefully fit and extract the parameters of the broad H$\beta$ emission line, which is present in all spectra. The PCA decomposition procedure focused on the accuracy of continuum levels and extended features and penalized models that overestimate peaky features such as narrow emission lines. This can result in residuals of the narrow features in the host galaxy-subtracted AGN spectra. To account for this and accurately measure the width of the broad emission, we fit multi-component emission features around the H$\beta$ emission line. 

\begin{figure}[h] 
\centering
\includegraphics[width=0.95\linewidth]{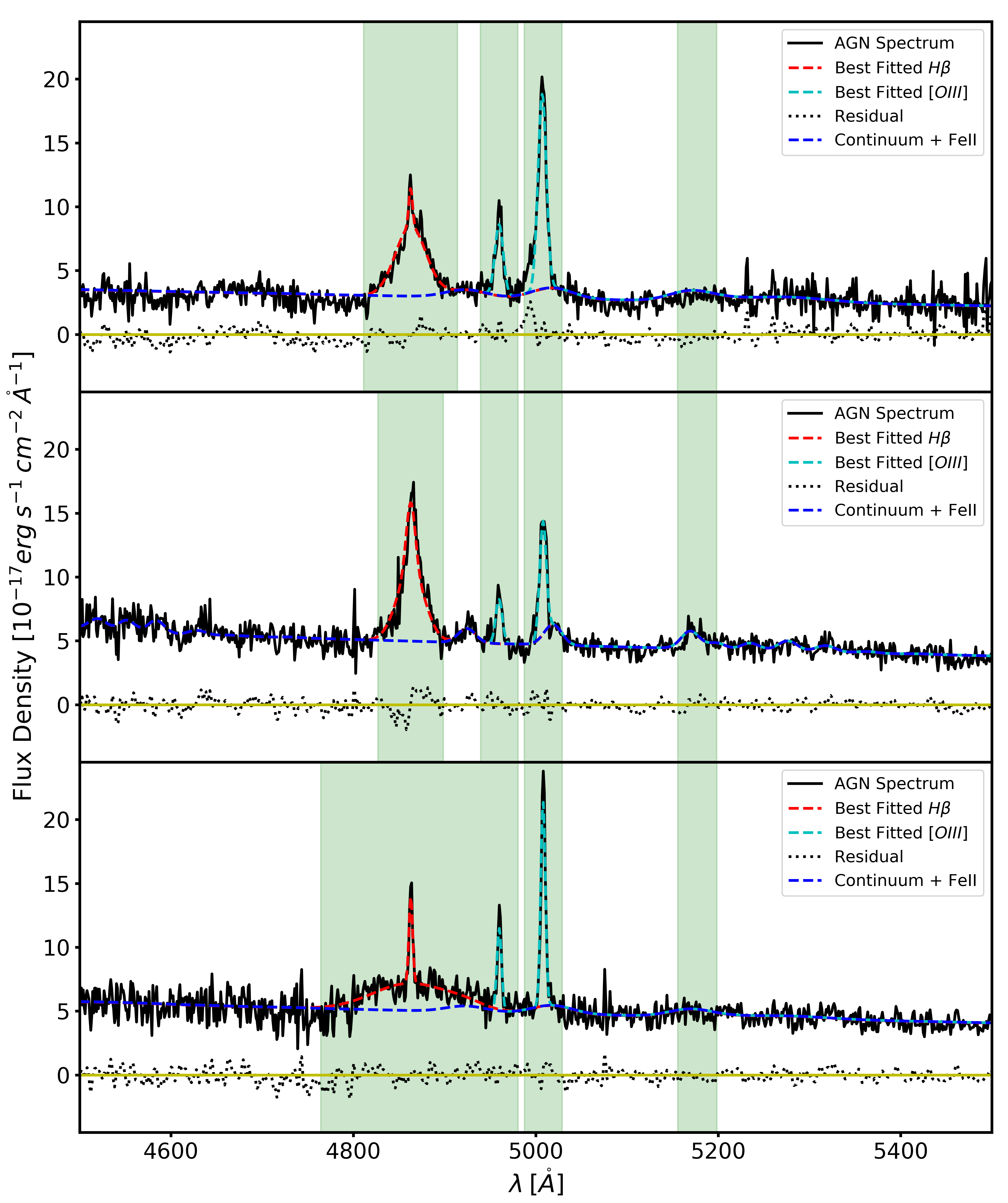}
\caption{Examples of the emission features fitting procedure and quality of spectrum modeling of the AGN components, showing 3 sources with the median signal-to-noise around H$\beta$ of 10. The fitted features include broad and narrow components of H$\beta$, [OIII] doublets, and $FeII$ emission features with continuum level with a slope as indicated by red, cyan, and blue dashed lines, respectively. The green-shaded regions are the masked areas of the spectra for the process of $FeII$ and continuum fits. The black-dotted lines represent the residuals after subtracting the best-fitted emission features and continuum.}
\label{figure9}
\end{figure}

In the fitting process, we shift the spectra to rest-frame and crop the wavelength range around 4,500 - 5,500 $\AA$. We start the process by fitting a broad (1,000 km s$^{-1}$ $<$ FWHM $\leq$ 10,000 km s$^{-1}$) and a narrow (FWHM $\leq$ 1,000 km s$^{-1}$) Gaussian profiles to the position of H$\beta$ and two Gaussian profiles (FWHM $\leq$ 10,000 km s$^{-1}$) at the positions of the [OIII] doublets, simultaneously. We then mask the fitted emission features with the width of $\pm$3$\sigma$ around the lines' centroids as shown in the green-shaded regions in Figure \ref{figure9}. It is well known that $Fe II$ emissions are significant and observable in the UV through the optical regimes of AGN spectra \citep[e.g.,][]{osterbrock1977, bronson1992, sulentic2000}. They overlap with the H$\beta$ and [OIII] features and can be modeled and fitted \citep[e.g.,][]{grandi1981, wills1985, bruhweiler2008, kovacevic2010, pandey2024}. Therefore, we fit the sum of five families of $FeII$ emission features with velocity width between 1,000-10,000 km s$^{-1}$ and a simple linear slope as the estimate of the continuum level through the unmasked patches of the spectrum \citep[e.g.,][]{oleas2016}. Then, we subtract the continuum and $FeII$ emission features from the original cropped rest-frame spectrum. After that, we fit the H$\beta$ and [OIII] features again and update the best-fitted parameters of the emission features. The fitting steps are repeated several times until the extracted parameters of the H$\beta$ emission features converge (i.e. changes in values less than 1\%). From the fitting results of 100 randomly selected spectra in our sample, we determine that a minimum number of 3 iterations is necessary to yield a high-quality fit. This is based on the convergence of the H$\beta$ emission width values, the residual of the flux density centered at zero with no clear systematic offsets, and the root-mean-square deviation being comparable to the square root of the flux density variance. We show examples of the emission lines fit in Figure \ref{figure9}. By accurately fitting and subtracting $FeII$, we prevent the systematic overestimation of the broad H$\beta$ line's width. We perform the fitting algorithm on all host galaxy-subtracted spectra in our sample to accurately determine the width of the H$\beta$ for our AGN. However, our fitting routine results in a surplus of sources with FWHM of the broad H$\beta$ emission equal to 1,000 km s$^{-1}$. This is due to the fitting constraint imposed on the minimum width for broad emission features. Thus, we make another cut to remove sources with the best-fitted FWHM of H$\beta$ with values less than 1,100 km s$^{-1}$. Therefore, we arrive at the final sample of 9,583 AGN (7,579 sources in Group I and 2004 sources in Group II).

\subsection{Properties of the Low-Luminosity AGN Sample}

\begin{figure}[h] 
\centering
\includegraphics[width=0.95\linewidth]{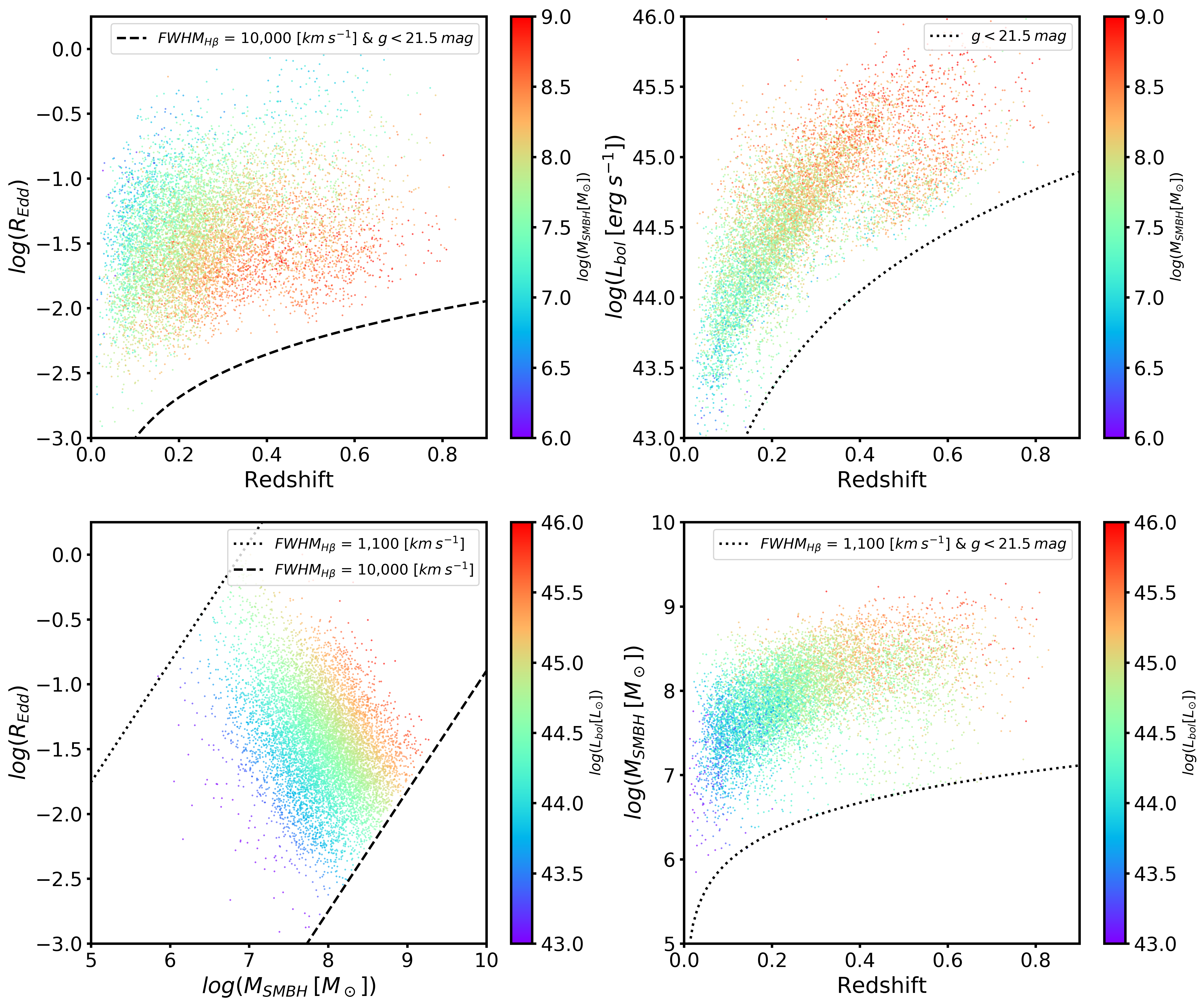}
\caption{Distribution of the properties of our low-luminosity AGN sample. {\bf Top panels:} The scatter along Eddington Ratio and bolometric luminosity versus redshift, from left to right. {\bf Bottom left:} The scatter along Eddington Ratio versus black hole mass. {\bf Bottom right:} The scatter along black hole mass versus redshift. The color scales on the top and bottom panels represent the black hole mass and bolometric luminosity, respectively.}\label{fig:sample_distribution}
\label{figure10}
\end{figure}

With the broad H$\beta$ line properly measured, we can compute the black hole mass for each source in our sample. We adopt the black hole mass equation from \citet{park15} using the width of the H$\beta$ line with the monochromatic continuum luminosity of the AGN measured at 5,100$\AA$ (i.e., log(M$_{SMBH}$/M$_\odot$)= 7.536 + 0.519log($\lambda$L$_{5100}$/10$^{44}$ erg s$^{-1}$) + 2log($\sigma_{H\beta}$/1,000 km s$^{-1}$)). The uncertainty in the determination of the host fraction (and AGN fraction) directly affects the measurements of flux density and ultimately the monochromatic luminosity at 5,100$\AA$. The reliability cut described and applied in Section \ref{sec:reliability} ensures that the uncertainty arising from PCA decomposition will be at most 25\% for $\lambda$L$_{5100}$ values. This yields an uncertainty of 0.07 dex in log(M$_{SMBH}$/M$_\odot$), which is marginal compared to the scatter due to the scaling relation itself of 0.4 dex \citep{park15}.

We can also calculate the bolometric luminosity adopting the bolometric corrections of \cite{duras20} using the rest-frame Bessel-B-band \citep{bessel1990} luminosity of the AGN (i.e., L$_{bol}$=5.18$\times$L$_B$). The B-band luminosities are computed by, first, convolving and integrating the extinction-corrected, host-galaxy-subtracted AGN spectra over the Bessel-B filter function. Next, the integrated B-band fluxes are multiplied with the redshift and distance dilution factors, (1+z) and (4$\pi$d$_L^2$); where d$_L$ is the luminosity distance. With these measurements, we can then determine the accretion rate as the ratio of the Eddington luminosity and the bolometric luminosity. Then, we compute the Eddington luminosity as L$_{Edd}$ = 1.2 x 10$^{38}\times$M$_{SMBH}$ where luminosity is in erg s$^{-1}$ and black hole mass is in solar units \citep{eddington}.

Figure \ref{figure10} shows the extracted parameters (i.e., accretion rate, black hole mass, and bolometric luminosity) versus redshift with the lower left panel showing the black hole mass vs the accretion rate as the Eddington Ratio, R$_{Edd}$ = L$_{bol}$/L$_{Edd}$. The dashed and dash-dotted lines indicate selection effects imposed on our sample due to survey depth and upper/lower limits placed on the fitted width of the H$\beta$ emission line. We compare these properties to those for quasars in SDSS Stripe 82 \citep{macleod2010}. Most notable is the difference in the bolometric luminosity of our sample, which has a median log(L$_{bol}$[erg s$^{-1}$]) of 44.6 compared to the quasar sample's median value of 46.2, which is almost 100 times more luminous. Our median black hole mass, log(M$_{SMBH}$ [M$_\odot$]), is 7.8 compared to 8.9 for the quasar sample. Finally, the median accretion rate of our sample is 5\% Eddington, with 20\% of the sample having accretion rates less than 1\% Eddington. The accretion rate for the typical quasar sample is about 10 times higher. This demonstrates the unique nature of our sample of SDSS AGN to investigate trends of variability in previously unstudied AGN parameter space.

\section{Ensemble Variability Characteristics} \label{sec:ensemble variability}

We compute the ensemble variability, V, using the following equation adopted from \citetalias{li2018}:

\begin{equation}
 V = \sqrt{\Delta m^{2}-2\sigma _{S/N}^{2}}
 \label{eq:var}
\end{equation}

\noindent where $\Delta$m represents the magnitude difference of each AGN and $\sigma$$_{S/N}$ is the photometric noise as a function of S/N derived from the control galaxy sample in Section \ref{sec:variability}. Our equation differs slightly from \citetalias{li2018} since we add a factor of 2 before the noise term, a necessary correction described in \citet{koz16}. This accounts for the fact that our noise is computed as a standard deviation value and, therefore, must be doubled to represent the $\Delta$m values of AGN, which vary both above and below the median value.

\subsection{Structure Function}\label{sec:SF}

\begin{figure}[h] 
\centering
\includegraphics[width=0.95\linewidth]{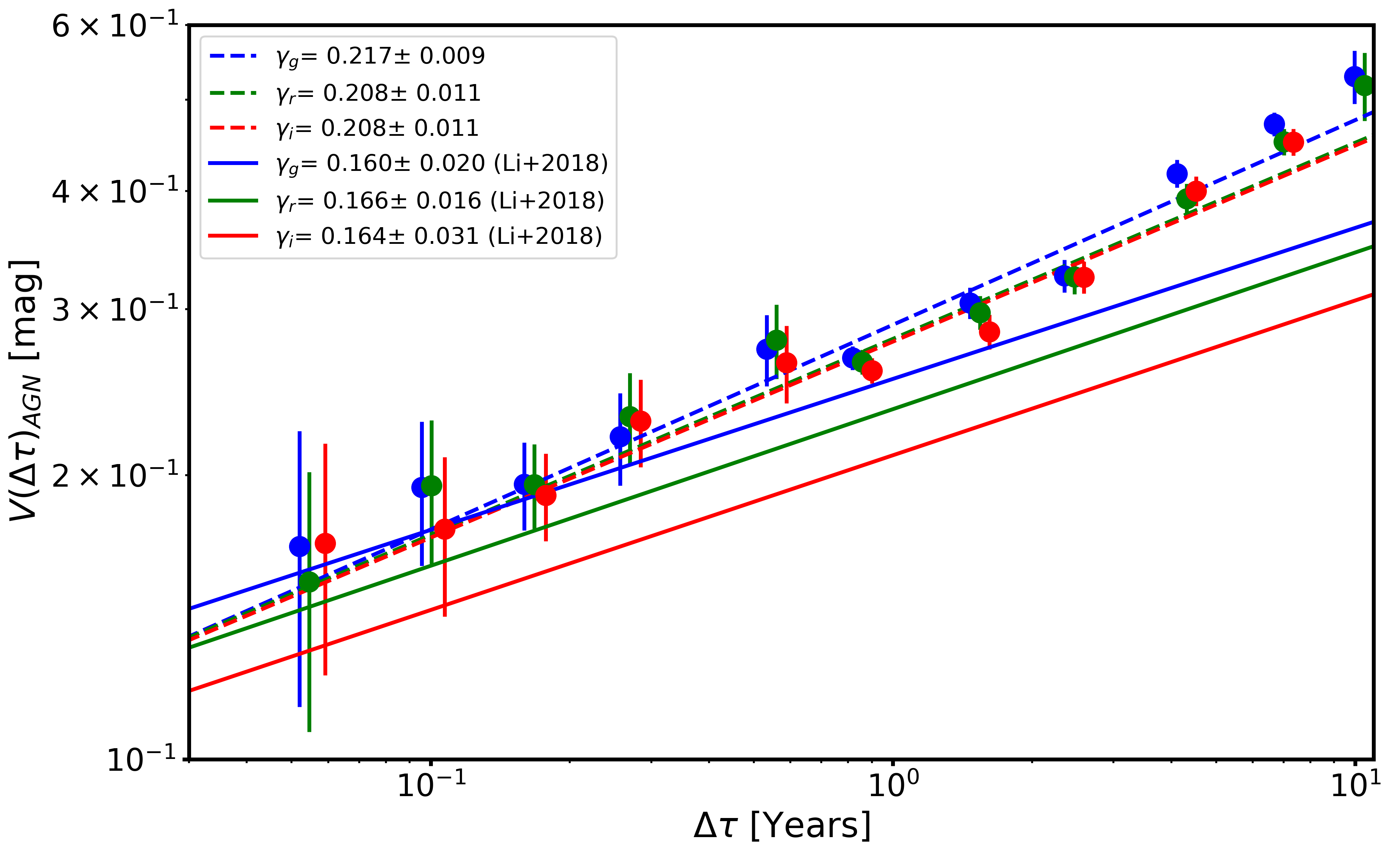}
\caption{The Structure Function (SF), or variability magnitudes as a function of time lags between two epochs. The blue, green, and red filled circles are the mean variability in each time lag bin with the error bars representing errors on the means of each bin in g-, r-, and i-bands, respectively. The dashed lines are the best-fitted SFs with the slope values and their bootstrapped uncertainties shown in the top-left legends. The solid lines are the Structure Functions of the quasars sample from \citetalias{li2018}.}\label{figure11}
\end{figure}

The Structure Function or SF is a method used to characterize the variability of AGN \citep[e.g.,][]{bauer2009, decicco2022, vandenberk2006, gs2014, li2018}. It is the relation between variability magnitude, V, and the rest-frame time lags, $\Delta\tau$, where the time interval between the two photometric measurements is corrected for time dilation effects by dividing by 1+z. Other techniques exist to quantify AGN variability such as computing the autocorrelation function (ACF). However, we compute the SF in order to compare with previously published results for ensemble variability of AGNs. In Figure \ref{figure11}, we show the ensemble SFs for the SDSS AGN sample, which is the averages of the V values of all AGN as defined in Equation \ref{eq:var} as a function of time-lag $\Delta\tau$. The error bars are computed as the error on the mean of each bin. 

The SF of each of the three filters is shown in blue (g-band), green (r-band), and red (i-band). The sample is binned into 12 equal intervals in log space (except for the first bin with the time lag of 0 to 10 days). Each rest-frame time lag bin contains between 62 and 1,618 sources with variability values above the photometric noises. We have applied a sigma-clipping routine to trim outliers from the sample in each interval. We perform 5$\sigma$-clipping with a maximum of 5 iterations to the $\Delta$m values in each bin to minimize the impact of individual outliers with large magnitude differences when computing the ensemble behavior of the AGN in each time lag bin. This process removes 1 to 2\% of sources from each bin or between 1 to 30 objects from the bins of 62 to 1618 objects. The outliers removed have unusually high variability (V$>$2 in some cases). Thus, this routine robustly minimizes the effect of these outliers in skewing the mean variability amplitude in each bin.



\begin{figure}[h] 
\centering
\includegraphics[width=0.95\linewidth]{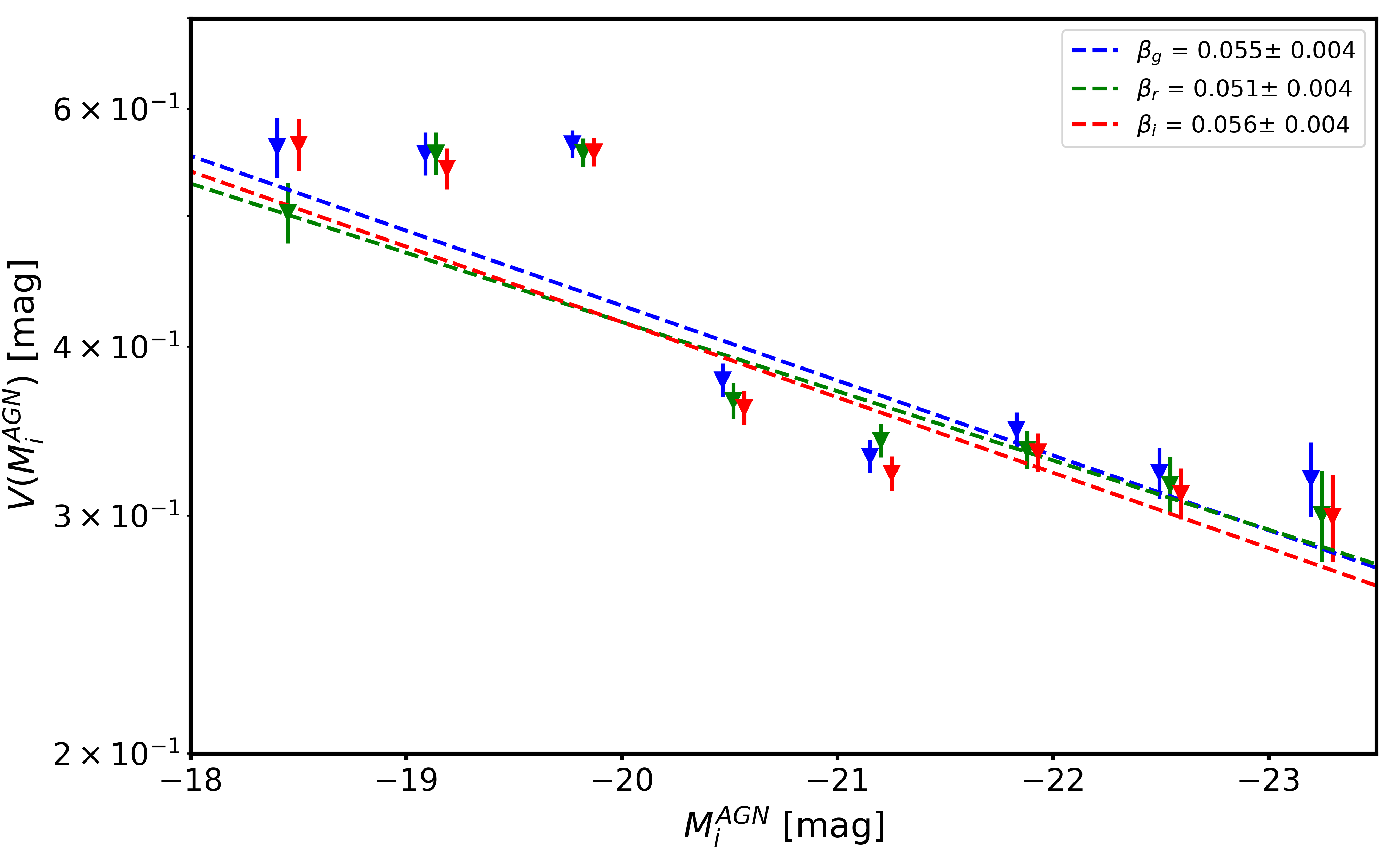}
\caption{The variability magnitudes as a function of absolute i-band magnitudes of the AGN. The blue, green, and red triangles are the mean variability in each magnitude bin with the error bars representing errors on the means of each bin in g-, r-, and i-bands, respectively. The dashed lines are the best-fitted models with the slope values and their bootstrapped uncertainties shown in the top-right legends.}\label{figure12}
\end{figure}

The AGN SF is generally found to increase with the rest-frame time lag, approaching a maximum variability amplitude, SF$_{\inf}$, between 0.1 and 0.5 at 1- to 3-year time intervals (see \citet{macleod2010}). We can parameterize the SF with the following equation, 

\begin{equation}
V = A(\frac{t}{365\: days})^{\gamma}
\label{eq:sf}
\end{equation} 

\noindent where A and $\gamma$ are the amplitude and slope of the function.  
As explained in \citetalias{li2018}, averaging over a large number of objects, which is necessary when computing ensemble statistics, will produce only a power-law relation since each source has its own set of SF parameters. Therefore, we fit only the slope of the relationship and show these fits as dashed colored lines corresponding to each band in Figure \ref{figure11}. The slopes of the SFs for our sample are between 0.208 and 0.217. These values are slightly shallower than those measured for low-luminosity AGN in \citetalias{gs2014} but agree within 1$\sigma$ uncertainties. Rather than determining the SF slope from least-square fitting, we perform 500 iterations of bootstrap resampling to quantify the spread of the SFs and adopt the medians and standard deviations as the slope values and their uncertainties as shown in Figure \ref{figure11}. Some studies have pointed out a caveat that estimating the uncertainty of the SF slope from least-square fitting may not represent the true underlying distribution of the slope values. For instance, \citet{emmanoulopoulos2010} have shown that the uncertainties on the slopes and breaks of blazars' SFs are several times underestimated when only considering the results of the least-square fit method.

Several previous studies have measured the slope of the ensemble Structure Function for quasars. We show the most recent ensemble SF for quasars from \citetalias{li2018} as solid lines in Figure \ref{figure11}. Colors correspond to the same bands as this work. The slopes for the quasar sample are generally shallower than our low-luminosity sources, with slopes closer to $\sim$0.160. Some of these differences in slopes could be because \citetalias{li2018} did not multiply the noise component by a factor of 2 before subtraction, which would have the effect of making the slopes shallower. An earlier study by \citet{vandenberk2004} found ensemble SF slopes for SDSS quasars that were much steeper, with values closer to $\sim$0.3 in all bands. This is more consistent with the values published by \citet{morg2014} who found the slopes to be $\sim$0.25. All of these slopes are shallower than the predicted power-law slope of 0.5 for the damped random walk (DRW) model often used to describe quasar variability \citep[e.g.,][]{kelly2009}. 

The overall amplitude of the low-luminosity sample is greater than that of the quasar samples, most notably at long intervals. We further explore the variability amplitude as a function of AGN luminosity in Figure \ref{figure12}. Here we determine the ensemble value of V in bins of AGN i-band absolute magnitude in rest-frame. We confirm the previously observed anti-correlation between AGN luminosity and variability amplitude \citep[e.g.,][]{zuo2012, simm2016}. We find that the increase extends to absolute magnitudes of -18. This is consistent with the results from \citetalias{gs2014} but is now shown with greater statistical significance. The slope of the relationship is shallower than that found in \citetalias{gs2014}. However, if we restrict to magnitudes fainter than -22, we find similar slopes and overall amplitudes with this previous study. 

There is a break in the variability amplitude at absolute magnitudes around -20. This appears to be caused by the fact that many of the intrinsically fainter AGN in our sample from Group I had the longest time intervals between their photometric and spectroscopic measurements due to observational survey parameters. For this reason, their overall variability amplitudes tend to be among the highest. The non-uniformity of the dataset highlights the importance of separating the variability properties and AGN characteristics to investigate these trends. In the next section, we focus on doing this to quantify relationships between variability and each AGN property while constraining the impact of the other characteristics.

\subsection{Variability versus Time lag, Luminosity, Redshift, Wavelength, Black Hole Mass, and Accretion rate}

We adopt the strategy of \citetalias{li2018} and perform a multidimensional fit to the SF as shown in Equation \ref{eq:sf}, with slope and variability amplitude as the dependences of bolometric luminosity, black hole mass, redshift, and rest-frame central wavelength of the observed band. In each parameter space, we split the sample into 5 bins, with equal spacing in log and linear spaces. For instance, the rest-frame wavelength and redshift spaces are each split into 5 equal-sized bins from $\lambda_c =$ 2,700 to 7,300 $\AA$, and from z = 0.01 to 0.84, respectively. While bolometric luminosity and black hole mass spaces are split into 5 equal-sized bins along the log-scale from L$_{bol}$ = 10$^{43}$ to 10$^{46}$ erg  s$^{-1}$, and from M$_{SMBH} $ = 10$^{6.5}$ to 10$^{9.5}$ M$_\odot$, respectively. Moreover, we have 10 bins in the log-scale for time lags from 0 to 10$^{3.7}$ days. This equates to splitting our sample into 6,250 cells across the 5 parameter spaces. Next, using one of SciPy's non-linear least squares optimizations (i.e., Levenberg-Marquardt algorithm), we fit the multidimensional parameters of the variability amplitude and slope following the formalism in the work by \citetalias{li2018} as expressed here:

\begin{equation}
\begin{split}
log(A) = log(A_0) + B_z log(1+z) + B_L log(\frac{L}{L_{46}}) + B_\lambda log(\frac{\lambda}{\lambda_4}) + B_M log(\frac{M}{M_9}) \\
\gamma = \gamma_0 + \beta_z z + \beta_L log(\frac{L}{L_{46}}) + \beta_\lambda log(\frac{\lambda}{\lambda_4}) + \beta_M log(\frac{M}{M_9}) \\
L_{46} = 10^{46}\: erg \: s^{-1}; \: \lambda_4 = 10^4 \AA; \: M_9 = 10^9 M_\odot\\
\end{split}
\label{eq:zLlM}
\end{equation}

\noindent where A$_0$ are the intrinsic level of variability amplitude at the time lag of 1 year. B$_z$, B$_L$, B$_\lambda$, and B$_M$ are the coefficients corresponding to the variability amplitude dependence on redshift, bolometric luminosity, the central rest-frame wavelength of the observed band, and black hole mass, respectively. The intrinsic slope of the Structure Function is represented by $\gamma_0$. While, $\beta_z$, $\beta_L$, $\beta_\lambda$, and $\beta_M$ are the coefficients corresponding to the SF slope dependence on redshift, bolometric luminosity, the central rest-frame wavelength of the observed band, and black hole mass, respectively.

To assess the uncertainties of these fitted parameters, we perform 500 iterations of bootstrap resampling of our AGN and disregard the cells with a total number of points less than 5. Thus, in any given iteration, the number of cells along any parameter space varies between 3 to 5, except for the time lag space. The multidimensional fitting is done following the prescription in Equation \ref{eq:zLlM}. The average and the $1\sigma$ uncertainty of each parameter is calculated based on the distribution of the best-fitted values of all bootstrap iterations and presented in Table \ref{tab:zLlM1} and \ref{tab:zLlM2}. 


The first three rows of Table \ref{tab:zLlM1} show the SF slope dependence on these parameters. We find very little significant correlations between the parameters and the SF slope. There is a positive 1$\sigma$ correlation with luminosity (Column 4), which is consistent with that identified in \citetalias{li2018}. Despite the broad range of masses sampled in our study, we find no significant dependence on black hole mass. The lack of dependencies found for redshift or wavelength may be due to the restricted redshift range of our sample. We check for changes in the fits when removing the redshift dependence in the second row and the wavelength dependence in the third row, and do not find any change in the correlations with the remaining parameters. 

The first three rows of Table \ref{tab:zLlM2} show parameter fits for the variability amplitude. We find that amplitude depends on bolometric luminosity with a 3$\sigma$ significance and black hole mass with 2.3$\sigma$ significance. The variability amplitude dependence on bolometric luminosity confirms the well-known negative correlation demonstrated in Figure \ref{figure12}. We find almost the same dependence on bolometric luminosity as that reported for the quasars in \citetalias{li2018}. This result verifies that the trend with bolometric luminosity is robust over more than 4 orders of magnitude, extending to fainter magnitudes than previously studied. We also test the removal of the redshift dependence in the second row of Table \ref{tab:zLlM2} and find that there is a negative correlation with wavelength with a 1.6$\sigma$ significance. As shown in Table \ref{tab:zLlM2}, removing both redshift and wavelength dependence does not impact the correlations with the other parameters significantly.

Although the significance is low, we find a positive correlation between variability amplitude and black hole mass. Previous studies have reported mixed findings concerning this correlation with some identifying a positive correlation \citep[e.g.,][]{wold2007, macleod2010, lu2019, suberlak2021} and others reporting negative or unclear results \citep[\citetalias{li2018};][]{zuo2012, simm2016}.  \citet{zuo2012} found that the dependence is positive when the influence of luminosity is excluded, but negative when the Eddington accretion rate is excluded. To test this, we revise the fitting prescription in Equation \ref{eq:zLlM} into the prescription as follows:

\begin{equation}
\begin{split}
log(A) = log(A_0) + B_z log(1+z) + B_{R_{Edd}} log(\frac{R_{Edd}}{R^{L_{46}}_{Edd}}) + B_\lambda log(\frac{\lambda}{\lambda_4}) + B_M log(\frac{M}{M_9}) \\
\gamma = \gamma_0 + \beta_z z + \beta_{R_{Edd}} log(\frac{R_{Edd}}{R^{L_{46}}_{Edd}}) + \beta_\lambda log(\frac{\lambda}{\lambda_4}) + \beta_M log(\frac{M}{M_9}) \\
{R_{Edd}^{L_{46}}} = \frac{10^{46}\: erg \: s^{-1}}{1.20\times10^{38} \frac{M_9}{M_\odot}}; \: \lambda_4 = 10^4 \AA; \: M_9 = 10^9 M_\odot
\end{split}
\label{eq:zRlM}
\end{equation}

\noindent where R$_{Edd}$ represents Eddington Ratio as an indicator of accretion rate onto a supermassive black hole and the other parameters remain the same as in Equation \ref{eq:zLlM}. This alternative formalism expressed in Equation \ref{eq:zRlM} replaces the bolometric luminosity term with Eddington Ratio, R$_{Edd}$. This change is also justified since the bolometric luminosity is related to black hole mass as inferred in Figure \ref{fig:sample_distribution}. Thus, the accretion rate may be a more fundamental AGN parameter. The results of this multidimensional fit are presented in lines 4 through 6 of Tables \ref{tab:zLlM1} and \ref{tab:zLlM2}. 

We find a negative correlation for accretion rate with variability amplitude with a 3$\sigma$ significance, indicating higher variability amplitudes for lower accretion rates. This is consistent with the results of previously published quasar studies \citep[e.g.,][]{zuo2012, simm2016, sanchez-saez2018, lu2019}, though this is the first time the trend is confirmed at accretion rates less than log(R$_{Edd}$) of -1.75 ($\sim$2\% Eddington). We also confirm the results of \citet{zuo2012} that the relationship with black hole mass switches from positive to negative depending on whether luminosity or accretion rate is considered. When the accretion rate is controlled, the trend becomes negative with 1.5$\sigma$ significance. The negative correlation between variability amplitude and wavelength is still observed with almost 2$\sigma$ significance, with bluer wavelengths displaying higher variability amplitudes. If we remove the redshift parameter from the fits, the relationship with wavelength becomes more significant, as does the relationship with accretion rate and amplitude, while the correlation with black hole mass becomes weaker. Removing both redshift and wavelength parameters does not significantly impact these results.

\begin{table}[h]
  \noindent 
  \centering
  \caption{Best-fitted multi-variable parameters to the slope of the Structure Function.}
  \begin{tabular}{cccccc}
    \hline
    \hline
    Fitted Parameters / Models  & $\gamma_0$ & $\beta_z$ & $\beta_L$ / $\beta_{R_{Edd}}$ & $\beta_\lambda$ & $\beta_M$ \\
    (1) & (2) & (3) & (4) & (5) & (6)\\
    \hline
    $\gamma (\gamma_0, z,L,\lambda, M)$     & $0.488\pm0.290$       & $-0.067\pm0.429$     & $0.150\pm0.143$  & $0.085\pm 0.349$       &  $0.010\pm 0.128$   \\
    $\gamma (\gamma_0,L,\lambda, M)$        & $0.328\pm0.126$       & --                   & $0.070\pm0.075$  & $-0.081\pm 0.235$      &  $0.042\pm 0.089$   \\
    $\gamma (\gamma_0,L, M)$                & $0.394\pm0.097$       & --                   & $0.090\pm0.079$  &  --                    &  $0.045\pm 0.097$   \\
 
    $\gamma (\gamma_0, z,R_{Edd},\lambda, M)$ & $0.462\pm0.316$   & $-0.025\pm0.493$      & $0.144\pm0.151$      & $0.025\pm 0.359$      &  $0.180\pm 0.146$   \\
    $\gamma (\gamma_0, R_{Edd},\lambda, M)$   & $0.328\pm0.134$   &  --                   & $0.079\pm0.089$      & $-0.127\pm 0.234$     &  $0.116\pm 0.072$   \\
    $\gamma (\gamma_0,R_{Edd}, M)$            & $0.378\pm0.105$   &  --                   & $0.084\pm0.099$      &  --                   &  $0.122\pm 0.089$   \\
    \hline
  \end{tabular}
  {{\bf Notes:} Each column of the table represents the following: (1) the parameters to be fitted across the column direction and the models used in the fitting process across the row direction; (2) the intrinsic slope of SF; (3) redshift coefficient; (4) bolometric luminosity or Eddington Ratio coefficient (depending on the fitted model); (5) rest-frame central wavelength coefficient; and (6) black hole mass coefficient.}
  \label{tab:zLlM1}
\end{table}

\begin{table}[h]
  \noindent 
  \centering
  \caption{Best-fitted multi-variable parameters to the Structure Function amplitude.}
  \begin{tabular}{cccccc}
    \hline
    \hline
    Fitted Parameters / Models  & $A_0$ & $B_z$ & $B_L$ / $B_{R_{Edd}}$& $B_\lambda$ & $B_M$ \\
        (1) & (2) & (3) & (4) & (5) & (6)\\
    \hline
    $A (A_0,z,L,\lambda, M)$       & $0.099\pm0.040$          & $1.075\pm 0.837$        & $-0.267\pm0.085$       & $-0.114\pm 0.151$         &  $0.139\pm 0.060$   \\
    $A (A_0,L,\lambda, M)$         & $0.186\pm0.024$          & --                      & $-0.188\pm0.045$       & $-0.156\pm 0.096$         &  $0.143\pm 0.045$   \\
    $A (A_0,L, M)$                 & $0.210\pm0.021$          & --                      & $-0.185\pm0.046$      & --                         &  $0.142\pm 0.055$   \\

    $A (A_0,z,R_{Edd},\lambda, M)$   & $0.103\pm 0.047$         & $1.052\pm1.007$        & $-0.258\pm0.087$       & $-0.125\pm 0.168$         &  $-0.123\pm 0.083$   \\
    $A (A_0,R_{Edd},\lambda, M)$     & $0.180\pm0.022$          & --                     & $-0.201\pm0.045$       & $-0.181\pm 0.098$         &  $-0.045\pm 0.026$   \\
    $A (A_0,R_{Edd}, M)$             & $0.211\pm0.024$          & --                     & $-0.187\pm0.058$       & --                        &  $-0.037\pm 0.042$   \\
    \hline
  \end{tabular}
      {{\bf Notes:} Each column of the table represents the following: (1) the parameters to be fitted across the column direction and the models used in the fitting process across the row direction; (2) the intrinsic variability amplitude of SF; (3) redshift coefficient; (4) bolometric luminosity or Eddington Ratio coefficient (depending on the fitted model); (5) rest-frame central wavelength coefficient; and (6) black hole mass coefficient.}

\label{tab:zLlM2}
\end{table}

\begin{figure}[h] 
\centering
\includegraphics[width=0.95\linewidth]{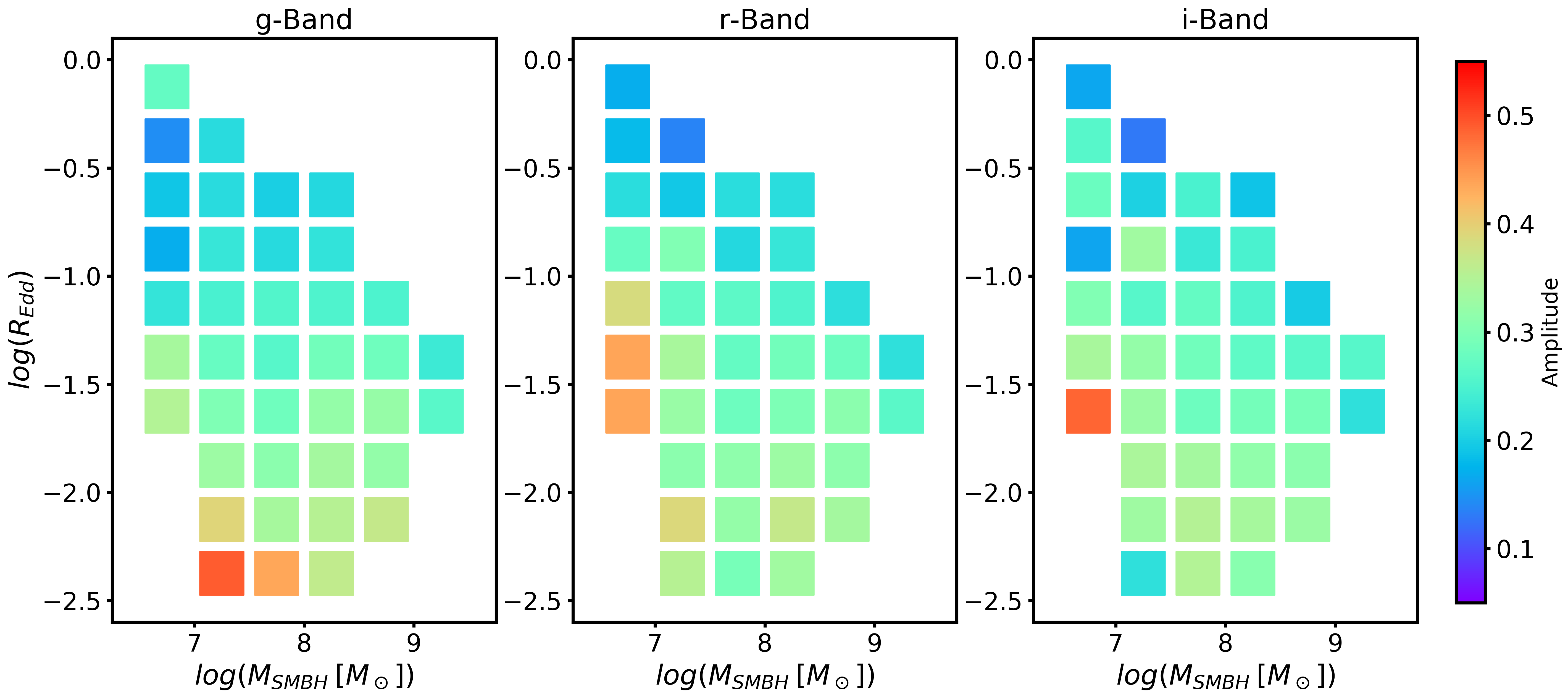}
\caption{The binned mean variability amplitudes of our AGN sample as a function of black hole mass and accretion rate as Eddington Ratio in g-, r-, and i-bands, respectively. The variability amplitude in each cell is represented by the color scale. The horizontal axis represents the log of the black hole mass in solar masses and the vertical axis represents the log of Eddington Ratio.}\label{figure13}
\end{figure}

Figure \ref{figure13} is a visual representation of the trends with variability amplitude, accretion rate, and black hole mass occurring in all three bands. The color of each square represents the average variability amplitude for all AGN in a given range of black hole masses and accretion rates. To obtain meaningful statistics, we exclude all the cells, which contain less than 10 AGN per cell. Warmer colors are associated with greater variability amplitudes. While it can be generally observed that the amplitude values increase with decreasing black hole mass, the amplitude increase is more obvious in the vertical direction as the accretion rate decreases. To test if the trends differ significantly between AGN populations with low and high accretion rates or between small and large black hole masses, we divide our sample at a black hole mass of log(M$_{SMBH} $) = 8 and accretion rate at log(R$_{Edd}$) = -1.25 and perform the multidimensional fit for variability amplitude with bootstrap resampling using the formalism in Equation \ref{eq:zRlM}. The direction of the trends remains the same for both accretion rate and black hole mass cases, but the negative correlation with black hole mass becomes greater (B$_M$ = -0.215$\pm$0.125) and slightly more significant among the low-mass black hole sub-sample. We also find that the negative correlation with accretion rate increases (B$_{R_{Edd}}$ = -0.307$\pm$0.126) but decreases in significance among the low-mass black hole sub-sample. 

As illustrated in Figure \ref{figure13}, the negative correlation between black hole mass and variability amplitude is primarily driven by the low accretion rate AGN and this is confirmed by our multidimensional fits. At high accretion rates, the change in variability amplitude with black hole mass disappears (B$_M$ = -0.004$\pm$0.185). However, at low accretion rates, the change in amplitude with black hole mass demonstrates a stronger correlation (B$_M$ = -0.162 $\pm$ 0.151) with slightly greater than 1$\sigma$ significance. Taken together, these results suggest that the accretion rate is the main driving parameter for variability amplitude with black hole mass having a less significant impact on this relation. 

To make certain that our sub-samples in black hole mass and accretion rate are not biased in terms of the SF time intervals sampled, we check for any dependence of these properties on $\Delta\tau$. We find that the average black hole mass remains between log(M$_{SMBH}$ [M$_\odot$]) = 7.9 to 8.1 as a function of $\Delta\tau$, with a standard deviation of $\pm$0.3, while the average Eddington Ratio clusters around log(R$_{Edd}$) = -1.40 to -1.60, with a standard deviation of $\pm$0.27. This ensures that our sub-samples represent the full range of parameters over all timescales.

\section{Discussion}\label{sec:discussion}

The significance of the accretion rate as a driving factor in variability amplitude that we identify has been previously suggested, though the precise reason behind the relationship is not entirely understood. \cite{macleod2010} argue that changes in the accretion rate are unlikely to produce this relationship since the accretion rate operates on longer viscous time scales that do not correspond to the shorter fluctuation time scales observed in variability studies. They also suggest that it could be the result of the amplitude anti-correlation with wavelength, which we also confirm in our study. Since higher accretion rates result in a hotter disk, this pushes the optical flux to a larger radius in the disk. If longer wavelengths are emitted further out in the disk, the lower variability amplitudes at long wavelengths could produce the observed relationship with the accretion rate. \citet{sun2020} proposed the CHAR (Corona-heated Accretion-disk Reprocessing) model, suggesting that the outer accretion disk and the innermost corona of an AGN are magnetically coupled. Thus, while the turbulence in the corona drives variability in the X-ray domain, it also fluctuates the heating rate of the accretion disk, giving rise to the variability in the UV/Optical domain. This model also points to the anti-correlations between variability amplitudes with black hole mass and the Eddington Ratio for timescale between days to years. This aspect of the CHAR model is consistent with our finding as shown in Table \ref{tab:zLlM2} and Figure \ref{figure13}. Similarly, another model has been proposed by \citet{yu2022} to explain the observed anti-correlation between amplitude and accretion rate. They argue that lower accretion rates result in a larger X-ray-emitting corona relative to the accretion disk. If the variable photons originate from this corona, this would produce a greater number of reprocessed photons in the disk resulting in a higher amplitude of variability. 

The \citet{yu2022} model also predicts the anti-correlation with black hole mass that we observe. They argue that the radius on the disk that emits at a fixed wavelength and accretion rate moves outward with increasing black hole mass. This results in a larger distance between the illuminating X-ray corona and the disk, producing lower intensity of the X-ray photons and thus a smaller amplitude of variability due to reprocessing of the photons in the disk. We observe the anti-correlation of variability amplitude with black hole mass when the accretion rate is accounted for. As predicted by their model, the black hole mass appears to have a secondary effect on the variability amplitude, displaying a weaker significance in the multidimensional fits than the accretion rate. 

Recently, \citet{arevalo2023} show that the relation between variability amplitude and black hole mass is negative when considering bins of constant accretion rate. They also show that the trend becomes more apparent on shorter timescales, arguing that the SF amplitude is relatively constant on long time scales and drops at shorter time scales, where the characteristic timescale varies with black hole mass. Our results are consistent with this finding, although we cannot confirm the trend being caused by changes in the characteristic time scale using ensemble statistics. We demonstrate, however, that the strength of the correlation is driven by sources at low accretion rates and is most apparent when sampling AGN with increasingly small black hole masses.

\section{Conclusions}\label{sec:conclusions}

We have presented an ensemble variability analysis of a sample of $\sim$9,600 AGN selected from the SDSS displaying broad emission lines and extended morphologies. Using principal component analysis, we decompose the host galaxy and AGN light to determine the AGN fraction for each source. Variability is determined using the difference between the photometric observations of each galaxy and the spectrophotometry in the g-, r-, and i-bands. Black hole masses are determined using the broad H$\beta$ emission line and flux at 5,100$\AA$.

Our AGN sample extends to fainter luminosities, smaller black hole masses, and lower accretion rates than previously studied samples, with the median value of black hole masses and accretion rates more than 10 times smaller than those of typical quasar samples used for variability analysis. Since variability properties have been shown to correlate with other AGN characteristics, the sample presented here allows us to probe these relationships in a previously unexplored parameter space to confirm trends or identify changes. We summarize our conclusions as follows:

\begin{enumerate}
\item{The slope of the Structure Function is similar to that observed for more luminous quasar samples but shallower than the predicted slope for the DRW model. There is a slightly significant positive correlation with the slope and luminosity, indicating that brighter AGN have steeper SF slopes.}

\item{We confirm the anti-correlation between variability amplitude and AGN luminosity with 3$\sigma$ significance. This trend continues to the faintest limit of our sample at log(L$_{bol}$ [erg s$^{-1}$]) $\sim$ 43.5 and appears robust over almost 4 orders of magnitude in luminosity.}

\item{We find that amplitude of variability is positively correlated with black hole mass when accounting for luminosity but negatively correlated when considering accretion rate in place of luminosity, indicating that less massive black holes display higher amplitudes of variability. The negative correlation is more pronounced among the lowest mass black holes and at the lowest accretion rates.}

\item{Our results suggest that the accretion rate is the dominant factor impacting variability amplitude. We confirm the anti-correlation for accretion rates as low as 1\% Eddington for the first time. We find that black hole mass has a secondary effect on the variability amplitude, with lower-mass black holes displaying more variability.}
\end{enumerate}

To better measure the variability properties of low-luminosity AGN, we plan to analyze the light curves for the SDSS AGN observed in recent variability surveys such as the Zwicky Transient Facility \citep{bellm2019}. Analyses of these light curves will allow us to test models of accretion physics and probe the limit where changes in the dominant accretion mode may occur. Understanding the variability nature of lower luminosity AGN will be important for interpreting observations with future large time-domain surveys such as the Rubin Observatory Legacy Survey of Space and Time (LSST; \cite{ivezic19}).

\section*{acknowledgments}

We thank Juan Oleas and Stephanie Podjed for their contributions to an earlier version of this work. We also thank Lei Hao for providing the code to fit AGN spectra with galaxy and quasar eigen spectra, which is the basis for our Python-based PCA decomposition routine.

KC thanks for the support of the National Astronomical Research Institute of Thailand (Public Organization) for access to computing facilities and workspaces. KC is also grateful for the funding support from Thailand Science Research and Innovation (TSRI) in the category of Fundamental Fund, project number 4692510 toward the completion of this study.

Funding for the Sloan Digital Sky Survey IV has been provided by the Alfred P. Sloan Foundation, the U.S. Department of Energy Office of Science, and the Participating Institutions. SDSS-IV acknowledges the support and resources from the Center for High Performance Computing at the University of Utah. The SDSS website is \href{https://www.sdss4.org/}{www.sdss4.org/}. SDSS-IV is managed by the Astrophysical Research Consortium for the Participating Institutions of the SDSS Collaboration including the Brazilian Participation Group, the Carnegie Institution for Science, Carnegie Mellon University, Center for Astrophysics | Harvard \& Smithsonian, the Chilean Participation Group, the French Participation Group, Instituto de Astrof\'isica de Canarias, The Johns Hopkins University, Kavli Institute for the Physics and Mathematics of the Universe (IPMU) / University of Tokyo, the Korean Participation Group, Lawrence Berkeley National Laboratory, Leibniz Institut f\"ur Astrophysik Potsdam (AIP),  Max-Planck-Institut f\"ur Astronomie (MPIA Heidelberg), Max-Planck-Institut f\"ur Astrophysik (MPA Garching), Max-Planck-Institut f\"ur Extraterrestrische Physik (MPE), National Astronomical Observatories of China, New Mexico State University, New York University, University of Notre Dame, Observat\'ario Nacional / MCTI, The Ohio State University, Pennsylvania State University, Shanghai Astronomical Observatory, United Kingdom Participation Group, Universidad Nacional Aut\'onoma de M\'exico, University of Arizona, University of Colorado Boulder, University of Oxford, University of Portsmouth, University of Utah, University of Virginia, University of Washington, University of Wisconsin, Vanderbilt University, and Yale University. 

Lastly, we would like to thank the anonymous referee, whose comments and constructive feedback contributed to the improvement of this paper. We also appreciate the time and effort that the referee and the editor have dedicated to the reviewing process of this manuscript.


\bibliography{reference}{}
\bibliographystyle{aasjournal}



\end{document}